%% file: mainConstitutionalConsensus_OPODIS.tex
\pdfoutput=1
\documentclass[acmsmall,nonacm,11pt]{acmart}
\usepackage[british]{babel}

\usepackage{array}
\usepackage{soul}
\usepackage{algorithm}
\usepackage{wrapfig}

\AtBeginDocument{\newtheorem{observation}[theorem]{Observation}}

\newcommand{\ia}{\textit{i}}
\newcommand{\ib}{\textit{ii}}
\newcommand{\ic}{\textit{iii}}
\newcommand{\iiv}{\textit{iv}}
\newcommand{\iv}{\textit{v}}

\newcommand{\calF}{\mathcal{F}}

\usepackage[shortlabels]{enumitem}
\setlist[itemize]{leftmargin=*}
\setlist[enumerate]{leftmargin=*}

\newcommand{\udi}[1]{\textcolor{blue}{[Udi: #1]}}

\newcommand{\andy}[1]{\textcolor{purple}{Andy: #1}}
\newcommand{\remove}[1]{}
\newcommand{\opodisCut}[1]{{}}

\newcommand{\temph}[1]{\emph{#1}}

\usepackage[most]{tcolorbox}

\newcommand{\mypara}[1]{\smallskip\noindent\textbf{#1.}}

\usepackage{xspace}
\newcommand{\ack}{\textsc{ack}\xspace}
\newcommand{\nack}{\textsc{nack}\xspace}
\newcommand{\inform}{\textsc{inform}\xspace}
\newcommand{\coronate}{\textsc{coronate}\xspace}

\setlist{nosep, leftmargin=*}
\setlist{itemsep=1pt, topsep=3pt, leftmargin=*}

\title{\smaller Constitutional Consensus for Democratic Governance}
\renewcommand{\shorttitle}{Constitutional Consensus}

\author{Idit Keidar}
\affiliation{%
  \institution{Technion}
  \country{Israel}}

\author{Andrew Lewis-Pye}
\affiliation{%
  \institution{London School of Economics and Political Science}
  \country{UK}}

\author{Ehud Shapiro}
\affiliation{%
  \institution{London School of Economics and Political Science}
  \country{UK}}
\affiliation{%
  \institution{Weizmann Institute of Science}
  \country{Israel}}

\author{Nimrod Talmon}
\affiliation{%
  \institution{Ben-Gurion University of the Negev}
  \country{Israel}}

\renewcommand{\shortauthors}{Keidar, Lewis-Pye, Shapiro, and Talmon}



\authorsaddresses{}

\begin{document}

\begin{abstract}
Permissionless-consensus-based Decentralised Autonomous Organisations (DAOs) are the prevailing paradigm for participant-governed digital organisations.  As participants have verified resources but no trusted identities, this ecosystem is necessarily plutocratic (one coin -- one vote). Here we offer, for the first time, a democratic (one person -- one vote) paradigm for the governance of digital communities and organisations, based on permissioned consensus and egalitarian decision processes.

In line with Lamport's vision of consensus as a self-governing parliament, in the democratic paradigm a constitution specifies both a decision making protocol as well as a consensus protocol, combined to let participants amend the constitution through constitutionally-valid decisions that are ratified by consensus.
To meaningfully instantiate this paradigm we integrate the disciplines of distributed computing and computational social choice, with the goal of providing a practical and efficient smartphone-based solution for the democratic self-governance of grassroots sovereign digital communities and organisations.  

The resulting Constitutional Consensus protocol employs (1) state-of-the-art Sybil-resilient democratic decision processes for amending the set of participants, supermajority threshold, and timeout; and (2) a novel Byzantine-fault tolerant consensus protocol that is DAG-based (following Cordial Miners) thus eschewing reliable broadcast, with dual-mode operation (following Morpheus) that is quiescent when idle,  has spontaneous leaders for isolated transactions, and formal round-robin leadership during high throughput. 
Constitutional Consensus operates in epochs, each governed by the prevailing constitution, with the next epoch commencing via the ratification of a constitutionally-valid constitutional-amendment decision. The protocol achieves consensus in $3\delta$ with $O(n)$ amortized communication complexity during high throughput. The single-epoch protocol is quite simple, requiring only 25 lines of pseudocode; the multi-epoch protocol is grassroots, as each member can initiate and participate in multiple independent instances that can form, operate and grow concurrently without coordination or dependence on global resources other than the net.
\end{abstract}

\setcounter{page}{0}

\maketitle
\newpage

\section{Introduction}

\mypara{Participant-governance of digital organisations} Permissionless-consensus-based Decentralised Autonomous Organisations (DAOs)~\cite{ethereum:dao,ding2023survey,hassan2021decentralized,talmon2023social} are the prevailing paradigm for realising participant-governed digital organisations. As participants have verified resources (e.g., staked tokens) but no trusted identities, this ecosystem is necessarily plutocratic: voting power is proportional to token holdings (one coin -- one vote). Here we offer, for the first time, a democratic (one person -- one vote) paradigm for the governance of digital communities and organisations, based on permissioned consensus and egalitarian decision processes.

\mypara{Constitutional self-governance} Reconfiguration---changing the set of participants in a consensus protocol---has been studied extensively~\cite{lamport10.1145/279227.279229part-time,lamport2001paxos,lamport2009vertical,aguilera2011dynamic,spiegelman2017dynamic,ongaro2014raft}. This literature addresses the mechanism of \emph{how} membership changes execute safely, but not \emph{who decides} on them: the decision authority is either left unspecified or delegated to an external entity. Moreover, prior work applies reconfiguration only to membership, leaving the rest of the protocol fixed.\footnote{ Tezos~\cite{goodman2014tezos} has no notion of membership but can self-modify the protocol via plutocratic governance.} 
In line with Lamport's vision of consensus as a self-governing parliament~\cite{lamport10.1145/279227.279229part-time}, we propose a democratic paradigm in which a \emph{constitution} specifies both a decision-making protocol and a consensus protocol, combined to let participants amend the constitution through constitutionally-valid decisions that are ratified by consensus.
To meaningfully instantiate this paradigm we integrate the disciplines of distributed computing~\cite{lynch1996distributed} and computational social choice~\cite{brandt2016handbook}, with the goal of providing a practical and efficient smartphone-based solution for the democratic self-governance of grassroots~\cite{shapiro2023grassrootsBA} sovereign digital communities and organisations. The resulting Constitutional Consensus protocol presented here employs (1)~state-of-the-art Sybil-resilient~\cite{shahaf2020genuine} democratic decision processes for amending the set of participants~\cite{bulteau2021aggregation}, supermajority threshold~\cite{abramowitz2021amend}, and timeout~\cite{shahaf2019sybil}; and (2)~a novel Byzantine-fault tolerant consensus protocol that is DAG-based (following Cordial Miners~\cite{keidar2023cordial}) thus eschewing reliable broadcast, with dual-mode operation (following Morpheus~\cite{lewis2025morpheus}) that is quiescent when idle, has spontaneous leaders for isolated transactions, and formal round-robin leadership during high throughput.

\mypara{Digital social contracts} A digital social contract~\cite{cardelli2020digital} is a voluntary agreement among a set of people, specified, fulfilled, and enforced digitally. Many digital social contracts --- including social graphs~\cite{shapiro2025atomic}, social networks~\cite{shapiro2023gsn}, and digital currencies~\cite{shapiro2024gc,lewis2023grassroots}---require no consensus at all. Those that do require consensus among the contract's participants---notably democratic governance and federation of sovereign communities~\cite{shapiro2025GF}---may execute the protocol on the participants' own devices. Constitutional Consensus is designed for such social contracts. Unlike smart contracts~\cite{de2021smart}, whose execution is enforced by global consensus among third-party validators on a blockchain, digital social contracts are among people with mutually-known identities and are enforced by the participants themselves. This distinction---summarised in Table~\ref{table:contracts} in Appendix~\ref{appendix:related-work}---has a direct architectural consequence: governance is democratic rather than plutocratic.

\mypara{Technical approach} Constitutional Consensus operates in epochs, each governed by the prevailing constitution, with the next epoch commencing via the ratification of a constitutionally-valid constitutional-amendment decision. The single-epoch protocol is quite simple, requiring only 25 lines of pseudocode. After GST, a leader block is finalised within $3\delta$, and the amortized communication complexity per transaction is $O(n)$ during high throughput.

\mypara{Grassroots} Because Constitutional Consensus requires no external authority, global ledger, or staking globally-controlled capital, the multi-epoch protocol is \emph{grassroots}~\cite{shapiro2023grassrootsBA,shapiro2025atomic}: each member can initiate and participate in multiple independent instances that can form, operate and grow concurrently without coordination or dependence on global resources other than the net. Any group of people can bootstrap their own instance on their smartphones. This is in contrast with Proof-of-Stake protocols like Ethereum 2.0~\cite{buterin2020combining}, where security depends on the staked token having real-world value, creating a winner-take-all dynamic that precludes independent instances.

\mypara{Paper structure}
Section \ref{section:model} provides the model and problem definitions.
Section \ref{section:protocol} presents single-epoch Constitutional Consensus, with no constitutional amendments.
Section \ref{section:democratic-processes} presents Sybil-resilient democratic decision processes for amending the set of participants, supermajority threshold, and timeout. 
Section \ref{section:constitution} presents the formal amendment mechanism and epoch transitions.
Section \ref{section:related} summarises related work, with a full survey in Appendix~\ref{appendix:related-work}.
Section \ref{section:conclusions} concludes. Correctness proofs and other technicalities are deferred to appendices.

\section{Model and Problem}
\label{section:model}

We assume a set of \emph{agents} $\Pi$, each has a unique key-pair and identified by its public key $p\in \Pi$.
While $\Pi$ is potentially-infinite, we refer only to finite subsets $P\subset \Pi$ of it.
 A \emph{correct} agent follows the protocol. An agent that does not follow the protocol is \emph{faulty} or \emph{Byzantine}.

We assume an \emph{eventual synchrony} (\emph{partial synchrony)} model in which every message sent between correct agents is eventually received (this abstracts a dissemination layer that can use acks, gossip, or network coding, and is orthogonal to the consensus protocol).  Message latency is unbounded, but in every run there is a time called the \emph{Global Stabilization Time (GST)}, which is unknown to the agents, after which all messages arrive within a least upper bound $\delta$ time for some $\delta >0$.  


In general, the constitution specifies the entire set of protocols employed---decision making processes as well as the consensus protocol for ratifying them.  In our case, these consist of known democratic decision processes recalled in Section~\ref{section:democratic-processes} and the novel Constitutional Consensus protocol, which is the main subject of this paper.  In this work the constitution assumes both of these protocols to be fixed except for three amendable components, the participant set, the supermajority threshold, and the timeout parameter.  

A protocol  proceeds through a sequence of \emph{epochs}, each commences with a new constitution, referred to as the \emph{prevailing constitution} during the epoch, and ends when the constitution is \emph{amended}, namely replaced with a new constitution:

\begin{definition}[Amendable component of the Constitution\opodisCut{, Participants, $\sigma$-Supermajority}]
\label{definition:constitution}
The \emph{amendable component of the constitution} is the triple $(P,\sigma,\Delta)$ where:
\begin{enumerate}
    \item $P\subset\Pi$ is  a set of agents, referred to as its \temph{participants}, with $n:=|P|$,
    \item $\frac{1}{2}\le \sigma < 1$ is a fraction that specifies a supermajority among the participants, 
    \item $\Delta>0$ is a presumed upper-bound on message delay after GST among the participants.
\end{enumerate}
We  may omit $\Delta$ when discussing protocol safety and refer to the constitution $(P,\sigma)$. 
A \temph{$\sigma$-supermajority}  among the participants is a fraction $Q\subseteq P$ such that $|Q| > \sigma n$.
\end{definition}
In the following  we may use \emph{the constitution} as a shorthand, e.g.  ``the prevailing constitution'' instead of ``the amendable component of the prevailing constitution''.

\remove{
\mypara{Agents vs Participants}
Throughout this paper, we distinguish between \emph{agents} and \emph{participants}: Any member of $\Pi$ is an \emph{agent}, while members of the set $P$ in a constitution $(P, \sigma, \Delta)$ are \emph{participants} in the epoch with this prevailing constitution. Only participants execute the protocol during an epoch, though any agent may become a participant through constitutional amendments.
}

\mypara{Supermajorities, Sybils and Byzantines} 
Our protocols are defined using a supermajority that aims to overcome participants that are (1) Byzantine (arbitrarily-behaving)~\cite{lamport1982byzantine}, which hamper consensus and (2)  Sybil (with fake or duplicate identity)~\cite{shahaf2020genuine}, which hamper equality.  Assuming there are at most $f<n$ such faulty participants, we define  $\sigma := \frac{n+f}{2n}$. 
Equivalently, $\sigma$ is chosen with the assumption that there are at most $f=(2\sigma-1)n$ faulty participants.   Both protocols ensure safety with any $f<n$ and liveness if $f < \frac{1}{3}n$.
\opodisCut{
For example, for $f=0$,  $\sigma = \frac{1}{2}$;
for $f=\frac{1}{2}n$, $\sigma = \frac{3}{4}$; 
and if everyone but one agent is faulty then $\sigma = \frac{2n-1}{2n} = 1-\frac{1}{2n}$, namely unanimity is required.  
For $n = 3f+1$ (the standard assumption),  $\sigma = \frac{n+\frac{n-1}{3}}{2n} = \frac{2}{3}-\frac{1}{6n}$, namely a supermajority greater-or-equal to $\frac{2}{3}n$ is required.

}
Typically, permissioned consensus protocols wait to hear from ``at least $n-f$ agents among $P$'', which in the standard case of $n = 3f+1$ (but not otherwise) happens to be equivalent to ``a $\sigma$-supermajority among $P$''. While Byzantines and Sybils are related (e.g.  Byzantines may aim to create and control multiple Sybils to amplify their ability to cause harm) they are not the same, and  a realisation of Constitutional Consensus may choose to incorporate a tighter supermajority for each, say $\sigma_B$ for consensus and $\sigma_S$ for democratic decisions; for simplicity, we employ here one supermajority $\sigma$ for both.

\mypara{Constitution safety and liveness} The (amendable component of the) constitution thus defines the protocol's assumptions about the environment: 
A constitution $(P,\sigma,\Delta)$ is \emph{safe} if 
$P$ is the set of agents running the protocol; there is a correct $\sigma$-supermajority in $P$, and every $\sigma$-supermajority in $P$ includes a majority of correct agents, 
hence the intersection of any two $\sigma$-supermajorities contains a correct agent. A constitution is  \emph{live} if $f < \frac{1}{3}n$ and after GST any message sent among correct participants arrives within $\Delta$.


\remove{
\begin{definition}[Safe and Live Constitution]\label{definition:constitution-safe-live}
A constitution $(P,\sigma,\Delta)$ is \temph{safe} if $P$ includes at most $f=(2\sigma-1)n$ faulty participants and \emph{live} if $f < \frac{1}{3}n$ and after GST any message sent among correct participants arrives within $\Delta$.
\end{definition}
}
Constitution safety and liveness attest to the validity of the assumption of the constitution regarding the participants and the network connecting them. 
Note that $\Delta$ bounds the message delay among \emph{correct} (i.e., participating and online) agents after GST, not the time until every agent comes online.  The dual-mode design 
means the protocol is quiescent when there are no pending transactions, so agents need not be continuously available.  When a transaction arrives, only a correct $\sigma$-supermajority must respond within $\Delta$; agents that are temporarily offline are accounted for by the choice of~$\sigma$.  

\mypara{Constitutional Consensus problem definition}
A constitutional consensus protocol may have multiple \emph{instances}. Each instance commences with an \emph{initial constitution} $(P,\sigma,\Delta)$\opodisCut{
that defines its participating agents $P$ and the key protocol parameters $\sigma$ and $\Delta$}. As in standard consensus,  participants spontaneously issue transactions (these are the inputs to the protocol) and every agent outputs a sequence of transactions. 

In addition, Constitutional Consensus assumes that participants engage in decision processes as specified by the constitution, which may result in decisions to amend the prevailing constitution.  As an example, Section~\ref{section:constitution} shows how known democratic decision processes can be applied in this new context.  Note that the parameters $P$ and $\sigma$ are shared: they govern both the consensus protocol and the democratic decision processes themselves.  The interface between  the Constitutional Consensus protocol and the constitutional democratic decision processes is as follows:
(\ia) a democratic process may produce constitutional amendment decisions, which become known to the participants once made; (\ib) participants can verify that a constitutional amendment decision is \emph{valid} according to the prevailing constitution, as specified in  Section~\ref{section:constitution}. A valid constitutional amendment decision defines a new constitution $(P',\sigma',\Delta')$\opodisCut{, in which the participants $P'$ and/or the parameters $\sigma'$ and $\Delta'$ are the result of amending the prevailing constitution}. An \emph{initial constitution} is the new constitution of a valid amendment of a constitution with an empty population. 

Constitutional amendment decisions are provided to the consensus protocol as inputs by participants that know of them.  Thus, the protocol's 
output sequence  includes both transactions and valid constitutional amendment decisions, as follows:

\begin{definition}[Constitutional sequence, epoch, epoch sequence, prevailing constitution, participants]
    A \temph{constitutional sequence} is a sequence $T = t_1, t_2, \dots$ of transactions and valid constitutional amendment decisions beginning with an initial constitution. Each constitutional amendment decision $d$ in the sequence defines a new \temph{epoch},  its \temph{epoch sequence} is a maximal contiguous subsequence of $T$ that begins with $d$ and does not include another such decision, and its \temph{prevailing constitution} is the new constitution of $d$. 
    An agent \temph{participates} in an epoch if it is a member of the population of its prevailing constitution.
\end{definition}

The constitutional output-sequences produced by correct agents executing a constitutional consensus protocol instance must satisfy the following safety and liveness conditions:

\begin{definition}[Constitutional Consensus Correctness]\label{definition:safety-liveness}
A constitutional consensus protocol is \temph{correct} if, in any run of it in which all valid constitutional amendments  are safe, there exists a constitutional sequence 
$T = t_1, t_2, \dots $ 
for which the following 
properties hold:
    \begin{description}
    \item[Consistency] Every correct agent's output sequence is a subsequence of $T$.
    \item[Validity] If a correct agent $p$ outputs $t\in T$, then $p$ is a participant in $t$'s epoch.  
    \item[Epoch Liveness] For a correct participant $p$ in an epoch $d$ in $T$,  $p$'s output  eventually includes all the transactions in $d$'s epoch sequence.  Furthermore, if $d$ is the last epoch in $T$, then every transaction input by a correct participant in epoch $d$ is included in $T$.
    \item [Constitution Amendment Liveness]
    If a valid constitutional amendment decision $d$ is provided as input to the protocol by a correct agent, then $T$ eventually includes $d$.
\end{description}

\end{definition}

Note that with no amendments, this reduces to the standard consensus correctness definition.  Progress is guaranteed after GST even under infinitely many amendments, though frequent amendments may starve normal transactions; in practice, amendments are expected to be rare.

\section{Single Epoch Constitutional Consensus}
\label{section:protocol}

Here we present the operation of the Constitutional Consensus protocol during a single epoch. We overview the protocol's operation, deferring its formal description and correctness proofs to Appendices \ref{section:formal} and \ref{section:safety-liveness}, resp. We begin (Section \ref{ssec:CC-safety}) by explaining the finalisation process, which ensures the protocol's safety.
We then discuss the conditions for generating new protocol blocks so as to guarantee liveness (Section \ref{ssec:CC-liveness}). 
Finally, Section \ref{proto} summarises the protocol's state, initialisation, and output.

\subsection{Safety: Finalising Transactions}
\label{ssec:CC-safety}

\mypara{Blocklace-based ordering}
The protocol is based on a \emph{blocklace}~\cite{keidar2023cordial}, a
cryptographic DAG data structure consisting of \emph{blocks}, each of which includes a \emph{payload} and hash-based \emph{pointers} to other (earlier) blocks. A block is signed by its creator; a block created by agent $p$  is called a $p$-\emph{block}.

An \emph{initial} block has no pointers.
The DAG structure defines a partial order on blocks: we say that a block $b$ \emph{observes} a block $b'$ in a blocklace $B$ if there is a path from $b$ to $b'$ in $B$. (In particular, every block observes itself). 
The \emph{depth} of a block $b$,  $\textit{depth}(b)$, is $0$ if $b$ is initial else $\textit{depth}(b)=\textit{depth}(b')+1$ where $b'$ is a maximal-depth block pointed to by $b$.  The \emph{closure} of a block $b$, denoted $[b]$, is the set of blocks observed by $b$. 

The protocol employs an evolving shared blocklace, of which each participant $p\in P$  maintains a local copy $B$. 
In each epoch, the protocol creates a  
blocklace that begins with a \emph{constitutional genesis block} $b$, whose payload specifies a new constitution $C=(P,\sigma,\Delta)$. 
Every participant commences the epoch with its copy of 
$B$ holding $\{b\}$.
During the epoch, participants create new blocks that observe their latest blocklace $B$ and  disseminate them to other participants. Blocks' payloads may include transactions the agents wish to order and also serve as protocol messages to drive agreement.

While a blocklace is only partially ordered, the protocol totally orders the blocks in an epoch (and the transactions included therein) in a consistent way, producing the same output sequence at all participants in an epoch.  Moreover, epochs are also totally ordered, so that the totality of all blocks of all epochs are also totally ordered, even though not always by the same participants.

\mypara{Waves, modes, and leaders} 
The blocklace is partitioned into a sequence of waves. For $k>0$, the $k^{th}$ wave in $B$
is the set of blocks of depths $3k-2$, $3k-1$, and $3k$ in $B$,  referred to as the the \emph{first}, \emph{second}, and \emph{third round} of the wave, respectively.
The $0^{th}$ wave of a blocklace is defined to be the singleton holding its constitutional genesis block.
%

The protocol has two operation modes, \emph{high} and \emph{low throughput}. In high throughput mode, agents continuously generate blocks in all rounds. To break the symmetry among agents, we use a 
deterministic (and fair) \emph{leader} function that defines a unique \emph{formal leader} $p_\ell  \in P$ in the first round of each wave. Within each wave, the protocol strives to finalise a first-round $p_\ell$-block (if it exists).

In low throughput mode, agents are quiescent as long as they have no new transactions to send. The protocol only quiesces at the end of a wave, after all previously sent transactions have been ordered. The quiescent state can last for seconds, minutes, or even hours, as in small social communities, transactions might be infrequent. Once any participant has a transaction to order, it initiates the next wave by sending a first round block. All other agents join this wave by issuing second, and later third, round blocks. In this scenario, the single first round block in the wave is considered a \emph{spontaneous leader}, and the protocol strives to finalise it. 

\mypara{Finalising blocks in waves}
Finalisation within a wave proceeds in three steps.
The first round of the wave includes candidates for finality, the second  includes endorsements of candidates, and the third  includes ratifications, as we now explain. 

But first, we need to disregard \emph{equivocations}, namely, two blocks by the same agent that do not observe each other.  We say that a block $b$ \emph{approves} a block $b'$ if $b$ observes $b'$ and does not observe any block equivocating with $b'$.  A second-round block $b$ \emph{endorses} at most one first-round block in its wave: if the preceding wave is quiescent, $b$ endorses the sole first-round block it approves (if unique); if non-quiescent, $b$ endorses the formal leader's block (if approved). A first-round block is \emph{ratified} by a third-round block that approves a $\sigma$-supermajority of endorsing second-round blocks, and \emph{final} once a $\sigma$-supermajority of third-round blocks ratify it. Since correct agents do not equivocate, at most one block per wave can be ratified or final.

\begin{figure}
\begin{minipage}[t]{.35\textwidth}
 \vskip 0mm
    \centering
    \includegraphics[width=\linewidth]{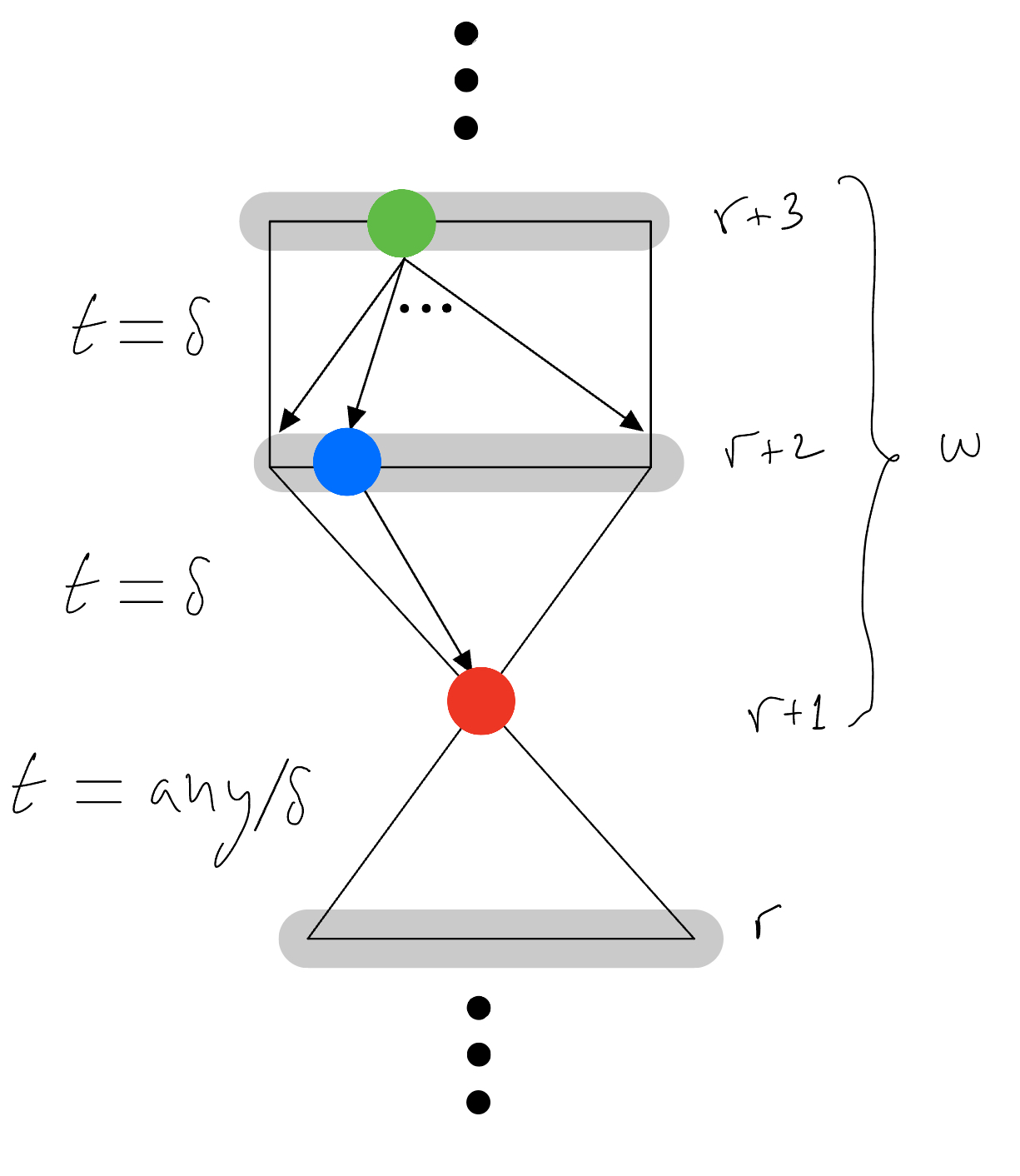}
\end{minipage}%
\hfill
\begin{minipage}[t]{.65\textwidth}
 \vskip 2mm
\small 
Round $r+1$ has a single leader block (red) approving a supermajority in round $r$ (the last round of the previous wave). Round $r+2$ consists of a supermajority of blocks (blue as an example) pointing only to (and thus endorsing) the first-round leader block. Round $r+3$ consists of  a supermajority of blocks (green as an example) each pointing to all  the second-round blocks, and thus ratifying the first-round leader block, which is thus finalised by the wave.
 \vskip 2mm
  
    \textbf{Low-throughput good case:} The  wave ending in round $r$ is quiescent, the spontaneous non-empty leader block (red) may be produced at any time after round $r$ has advanced. 
 \vskip 2mm
    
     \textbf{High-throughput good case:} The formal leader produces a block (red) within $\delta$ after the previous round $r$ is advanced.
\end{minipage}
 \vskip -2mm
    \caption{ \textbf{A good-case wave $w$.}  
     } 
    \label{fig:wave}
\end{figure}

\remove{
During low-throughput all waves are quiescent.  A wave commences when some agent has a new transaction. This agent issues a (first-round) non-empty block, and all agents follow promptly with two rounds of empty blocks: Second-round blocks that endorse the first-round block, and third-round blocks each approving all second-round blocks, upon which the first-round block becomes final. 

During high-throughput, no wave is quiescent, and the good-case scenario unrolls as above except that the first-round block is issued promptly by the wave's (formal) leader, whether or not it has any payload, and second- and third-round blocks are typically non-empty, with transactions to be ordered by the the next final leader block. 
}

We are now ready to define quiescence; the definition is mutually recursive with the definitions of endorsement, ratification, and finality. For the base, the $0^{th}$-wave is \emph{quiescent} and its constitutional genesis block \emph{final}. A subsequent wave is \emph{quiescent} if it includes a final leader block $b$, and all the blocks in the wave except $b$ have empty payloads. Thus, the protocol begins quiescent and remains in low throughput mode as long as there is at most one agent issuing transactions in each wave. It 
switches to high throughput mode whenever multiple agents send transactions in the same wave.
 
\mypara{Ordering}
As in other DAG-based protocols \cite{keidar2021need, keidar2023cordial, lewis2025morpheus}, once a leader block $b$ is finalised, the protocol deterministically orders all the transactions approved by $b$ that have not yet been totally ordered, and outputs the resulting sequence. Thus, as long as new blocks are being finalised, all the transactions in the system become ordered.

Ordering must be consistent at all agents. To this end, we leverage the property that at most one block is ratified or final in a wave. We define a recursive ordering function $\tau$. 
When called for a  block $b$, $\tau$ locates the most recent ratified and yet unordered block in $[b]\setminus \{b\}$ and recursively calls $\tau$ for this block. Once no such block exists, $\tau$ topologically sorts all unordered blocks approved by $b$.

\subsection{Liveness: Dissemination and Round Progress Rules}  
\label{ssec:CC-liveness}

In the previous section, we explained how transactions are finalised in the good case, when a first-round leader block is observed by a super-majority of endorsements in the second round and a super-majority of ratifications in the third.

We now discuss how we continue to progress through rounds when the good case does not occur, while taking care not to rush to the next round before giving the leader a chance to be finalised in case we are after GST.

To formalize these progress rules, we first define a round $r$ as \emph{advanced} in a blocklace $B$ if $B$ includes 
enough blocks to allow round $r+1$ blocks to be added to the blocklace safely. 

In second and third rounds, all correct participants send blocks, and these rounds advance once they include a super-majority of blocks. In a first round, only a leader (formal or spontaneous) sends a block in the good case. Hence a round is advanced once it includes a leader block. 
Going beyond the good case,  if during high-throughput agents fail to receive a first-round leader block in a timely manner, they issue first-round blocks of their own. 
Once the blocklace includes a super-majority of round-one blocks,
the round is advanced despite not including a leader block.

A second challenge is partial dissemination, whereby a faulty agent sends a block to some participants and not others, causing missing dependencies and potentially preventing some correct agents---including the formal leader---from advancing to the next round.

Partial dissemination is handled using two special \emph{dissemination-inducing} blocks, which trigger block forwarding and are not added to the blocklace: a {\sc nack}-block for missing dependencies and an {\sc inform}-block for an unresponsive leader.  

Before sending dissemination-inducing blocks, we wait sufficiently long for messages from correct participants to arrive after GST, so as not to induce extra traffic in the good case. Similarly, we have to carefully set the timeout for sending non-leader first-round blocks so as to give a correct leader sufficient time to make up any missing blocks and send its new block. The delays used in the protocol are explained in Appendix \ref{appendix:single-epoch-walkthrough}.

\subsection{Single-Epoch Constitutional Consensus Pseudocode} \label{proto}

In the Constitutional Consensus protocol, each agent $p$ maintains two sets of blocks: An input buffer $D$ and a local blocklace $B$. An epoch identifier $e$  points to the epoch's constitutional genesis block.   $B^e$  denotes the \emph{blocklace of epoch $e$ in $B$}.
Algorithm~\ref{alg:consensus} (Appendix~\ref{appendix:pseudocode-walkthrough}) presents the single-epoch Constitutional Consensus pseudocode, in which the prevailing constitution is fixed throughout the run. 
We use the terminology \emph{send/issue once} as a shorthand for send/issue if the block has not been sent/issued yet.

The protocol makes judicious use of delays, to ensure both liveness and efficiency. The rationale for the various protocol delays is given in Appendix~\ref{appendix:single-epoch-walkthrough}.

\mypara{Initialisation and input} The first epoch is initialised with 
$D \leftarrow \phi$, $B \leftarrow \{b\}$,  $e\gets b$,
where $b$ is a constitutional genesis block and $\emph{payload} \leftarrow \bot$. As input transactions arrive, they are added to \emph{payload}.
\opodisCut{
The initialisation of subsequent epochs is discussed in Algorithm~\ref{alg:ca}.
}
{\bf Output:}
Once the protocol finalises a block $b$, it orders the unordered blocks approved by $b$ using a recursive function $\tau$ defined in
Appendix \ref{appendix:tau}, and outputs their transactions.

\mypara{Latency and complexity}
After GST and in the good case, a leader block is finalised within $3\delta$. The amortized communication complexity per transaction is $O(n)$ during high throughput; details are in Appendix~\ref{appendix:latency-complexity}.

\input{democratic_processes}

\section{Constitutional Consensus with Constitutional Amendments}
\label{section:constitution}

We now extend the consensus protocol with constitutional amendments, which generalise reconfiguration in both scope (any protocol parameter, not just membership) and authority (decided by participants through constitutionally-specified processes, not by a leader or external entity); a  comparison with reconfiguration work is given in Appendix~\ref{appendix:related-work}. 
The protocol's correctness is proved in Appendix \ref{section:safety-constitutional}, and it being  grassroots~\cite{shapiro2023grassrootsBA}  in Appendix \ref{section:grassroots-is-grassroots}.

\remove{
In this work, the amendable component of the constitution consists of $(P,\sigma,\Delta)$ (Definition \ref{definition:constitution}), parameters which affect the safety, liveness, and performance of almost any consensus protocol for eventual synchrony.
}

%

The democratic decision processes described in Section \ref{section:democratic-processes} are part of the constitution: they specify how each amendable component may be changed.  The result of the \emph{amendment protocol} described above is a \emph{constitutional amendment decision}:

\begin{definition}[Constitutional Amendment Decision]
A \temph{constitutional amendment decision} is a pair $d=(h,x)$, where  $x=(\textsc{amend},\textit{Id},i,C,C')$, \textit{Id} is an \temph{instance identifier}, $i\ge 1$ is the \temph{epoch}  within this instance, $C,C'$ are constitutions, referred to as the \temph{old} and \temph{new} constitutions of $d$, resp., and $h$ is a set of signed votes. 
If $i=1$ then $C=\bot$ and $P_1$ are called the \temph{founders} of the protocol instance \textit{Id}---they are the signatories of its initial constitution  and the participants of the first epoch.

The decision $d$ is \temph{valid} in epoch $i$ of instance $Id$
with prevailing constitution $C=(P,\sigma,\Delta)$ if 
(1) $h$ consists of  a $\sigma$-supermajority in $P$ of signed votes for epoch $i$ in instance $Id$, (2) these are the first $\sigma$-supermajority of signed votes ordered in epoch $i$ of the constitutional consensus protocol instance $Id$, and (3) $C'$ is the result of applying the  amendment protocol to $h$ with prevailing constitution $C$. 
\end{definition}

Thanks to the use of consensus to order the votes and the democratic decision process, there is a unique valid amendment decision in each epoch of each instance of the protocol.

\opodisCut{
Note that the structure of $d$ is that of a constitutional genesis block. During a run of a protocol instance \textit{Id}, agents produce a possibly-infinite ordered sequence of constitutional amendment decisions $d_1,d_2,\ldots$ using democratic decision-making processes discussed above or others, where $d_i=(h_i,x_i)$ and $x_i=(\textsc{amend},\textit{Id},i,C_{i-1},C_i)$.   The  protocol commences with $d_1$ as the constitutional genesis block of the first epoch, and with each subsequent $d_i$ serving as the constitutional genesis block of the $i^{th}$ epoch.
}

\opodisCut{
\begin{proof}
By a counting argument, for this to happen a correct agent would have to sign two conflicting constitutional amendments, a contradiction.
\end{proof}
Note that while the safety depends only on $\sigma_i$ and $P_i$,  validity depends also on 
$\sigma_{i-1}$ and $P_{i-1}$, in order to adhere to the $h$-rule.
}


Constitutional amendments cause epoch transitions, which proceed in three phases. {\bf Amendment finalisation:} 
Once the amendment protocol at a participant $p$ results in a valid amendment $d$, $p$ includes $d$ as payload in every block it issues until $d$ is finalised. 
{\bf End epoch:} upon finalising a constitutional block with $d$, an agent produces a \coronate block, sends it to both old and new populations, and ceases participation; it must still respond to belated \nack requests from agents finalising the old epoch. {\bf Start epoch:} a member of the new population commences the new epoch upon receiving a $\sigma$-supermajority of \coronate blocks (and having sent one, if also in the old population). The full pseudocode is given in Algorithm~\ref{alg:ca} (Appendix~\ref{appendix:amendment-pseudocode}).
Liveness of constitutional amendments is guaranteed after GST since agents periodically send sets of votes, which after GST all become ordered. 

\opodisCut{
In the  good case, a new constitutional amendment $d$ is issued by the leader of a wave and finalised within that same wave. In the worst case, it might take arbitrarily many waves to finalise $d$ prior to GST or due to faulty leaders. 
After GST, a constitutional amendment block issued  by a correct leader among $P$ is bound to be finalised (and such a leader is bound to be selected eventually due to the fairness of $\textit{leader}_P$).  Once finalised, correct members of $P$ will issue a {\sc coronate} block and correct members of $P'$ will receive it, allowing the current epoch to end and the next one to begin. This argument informally establishes the liveness of the transition among constitutions due to constitutional amendments.
}

\section{Related Work}\label{section:related}

A full survey of related work is provided in Appendix~\ref{appendix:related-work}, organised into five areas.  It reviews smart contracts and digital social contracts~\cite{de2021smart,goodman2014tezos,cardelli2020digital}; DAOs and blockchain governance, including self-amending blockchains~\cite{hassan2021decentralized,wood2016polkadot,cip1694,kwon2016cosmos,camenisch2022ic}, DAO frameworks~\cite{cuende2017aragon,leshner2020compound,openzeppelin2021governor,snapshot2020,optimism2022governance}, critiques of plutocratic governance~\cite{schneider2022cryptoeconomics,siddarth2020watches,talmon2023social,buterin2021coinvoting}, and consortium platforms~\cite{Androulaki2018Fabric,Brown2016Corda}; reconfiguration of consensus protocols~\cite{lamport10.1145/279227.279229part-time,lamport2010reconfiguring,lamport2009vertical,aguilera2011dynamic,spiegelman2017dynamic,lorch2006smart,ongaro2014raft,bessani2014bftsmart,duan2022foundations,komatovic2025permissioned,Vizier2020ComChain,Bessani2020FromByzantine}; BFT consensus protocols, both leader-based~\cite{castro1999pbft,yin2019hotstuff,buchman2016tendermint} and DAG-based~\cite{keidar2021need,danezis2021narwhal,giridharan2022bullshark,gelashvili2022jolteon,babel2023mysticeti,raikwar2024sokdag}, including the two protocols on which Constitutional Consensus builds~\cite{keidar2023cordial,lewis2025morpheus}; and grassroots protocols and democratic identity, including centralised democratic platforms~\cite{barandiaran2024decidim,loomio2024,royo2020decidemadrid,consul2025,behrens2018liquidfeedback} and proof-of-personhood systems~\cite{ford2020identity,borge2017proof,kavazi2023humanode,idena2018,brenzikofer2019encointer}.  Constitutional Consensus combines three properties---democratic one-person-one-vote governance, decentralised BFT consensus, and self-governing constitutional amendment---and to the best of our knowledge, no existing system achieves all three.

\section{Conclusions}\label{section:conclusions}

The Constitutional Consensus protocol was devised to serve the needs of future grassroots platforms that operate digital social contracts.  A key design principle is that governance is democratic: participants themselves decide on constitutional amendments---who participates, the supermajority threshold, and the timeout---through decision processes grounded in computational social choice.  Initially, we expect low-throughput applications to prevail, as small digital communities engaged in constitutional democratic conduct use the platform to grow and evolve.  As communities federate to form ever-larger communities, and local community banks federate to form regional and national banks, higher-throughput execution would follow. Until then, simulating the various scenarios would be productive future work. Another direction is making the democratic decision-making process an amendable component of the constitution, making the system reflexive: the amendment process would then be subject to amendment through the very process it specifies.

\bibliographystyle{ACM-Reference-Format}
\bibliography{bib}

\newpage
\appendix

\input{formal_definitions}
\input{pseudocode_walkthrough}

\input{latency_complexity}

\input{single_epoch_delays_walkthrough}

\input{single-epoch-proof}
\input{amendment_setting}
\input{amendment_pseudocode}

\input{amendment-proof}

\input{grassroots}

\input{related_work_appendix}

\end{document}

%% file: democratic_processes.tex
\section{Sybil-Resilient Democratic Decision Processes}\label{section:democratic-processes}

The question of how a democratic constitution can be amended, democratically, is deep and fundamental~\cite{buchanan1965calculus}.
Here we develop democratic decision rules for amending each component of the constitution $(P,\sigma,\Delta)$, grounding our approach in the theory of computational social choice~\cite{brandt2016handbook}, using state-of-the-art machinery that fits to our setting, which could be summarised by the following three properties:
\begin{enumerate}
    \item \emph{Reality-aware}~\cite{shapiro2017reality}: The only type of decision is to amend the status quo; in particular, to amend the prevailing constitution~\cite{abramowitz2021beginning}.  Thus, the prevailing constitution $(P,\sigma,\Delta)$ is an ever-present reality: it governs the running consensus protocol regardless of whether anyone wishes to change it, with amendment rules treating it as the default and move away from it only when there is sufficient support.
    \item \emph{Sybil-resilient}~\cite{shahaf2019sybil,meir2024safe}: Change of the status-quo can be effected only by support large enough to necessarily reflect the will of the non-Sybil agents.\footnote{Sybil-resilient admission protocols~\cite{poupko2021building,shahaf2020genuine} during community formation are complementary to this work.}
    \item \emph{Aggregation over metric spaces with single-peaked domains}~\cite{bulteau2021aggregation}: We assume that: (1) each agent has an ideal point in each of the decision domains; (2) the domain has a metric, so that an agent prefers any point closer to their ideal point over any point further away from it; and (3) there exists an aggregation function that, given the ideal points of agents, produces a point that corresponds to the agent opinions.  This allows unifying proposing and voting:  Agents just state  their ideal points and the aggregation function identifies a most supported point; if those that prefer this point over the status-quo are a supermajority, it becomes the new decision.
\end{enumerate}

These properties allow for \emph{unified proposing and voting}, 
(see Appendix~\ref{appendix:amendment-setting}).
This means that participants only need only agree on a set of votes, and then each agent can independently apply the amendment rules (described below) to determine the next constitution.

\mypara{Constitutional amendment by consensus}
At a fixed periodic deadline (e.g., every Tuesday at midnight GMT),\footnote{Real-world realizations may wish to add response mechanisms for emergency situations.} participants use the consensus protocol to disseminate  votes: At the periodic 
deadline, each agent $p_j$ creates a transaction $T_j$ holding a set of signed votes including the latest vote it knows from each participant. The agent proposes this transaction as input to the single-epoch consensus protocol.

Once ordered, these transactions are not output to users but instead provided as input to the \emph{amendment protocol}, run independently by each agent. Once the amendment protocol gets, for the first time, a $\sigma$-supermajority of such ordered transactions $T_j$ in an epoch $i$ of an instance of the consensus protocol, it creates  a set holding the latest vote from each participant included in any of the received sets. The amendment protocol determines the next constitution based on this set and produces a \emph{constitutional amendment decision}. Note that the amendment protocol produces at most one amendment per protocol epoch.


\mypara{Amendment rules}\label{subsec:rules}
We now specify the concrete amendment rules 
given the agreed upon set of votes and the prevailing $(P,\sigma,\Delta)$; the updated constitution $(P',\sigma',\Delta')$ takes effect when the consensus protocol switches to a new epoch, as described in Section \ref{section:constitution}.


\mypara{(1) Amending $P$}
Each agent $i \in P$ votes independently on each potential participant $p$: yes (in) or no (out).  A candidate $p$ is added to (or retained in) $P'$ if and only if a $\sigma$-supermajority of the prevailing $P$ votes yes.  Additionally, every new member in $P' \setminus P$ must consent to joining, to prevent deadlock from non-responsive members~\cite{lamport10.1145/279227.279229part-time}.

This per-member $\sigma$-supermajority rule follows directly from the Sybil-resilient setting of Shahaf et al.~\cite{shahaf2019sybil} (for the setting of voting in a binary domain-in/out), who show that it is safe with respect to honest agents: a minority of Sybils cannot force membership changes.


\mypara{(2) Amending $\sigma$}
Each agent $i \in P$ states its preferred $\sigma_i \in [0.5, 1)$.  The value of $\sigma$ is updated according to the \emph{$h$-rule for constitutional amendment}~\cite{abramowitz2021beginning} (named after the $h$-index of citation metrics), as follows:
%
    (1) \emph{Raising $\sigma$:}  $\sigma$ is increased to the maximal $\sigma' > \sigma$ for which a $\sigma'$-supermajority of $P$ voted for a value $\geq \sigma'$.
    (2) \emph{Lowering $\sigma$:}  $\sigma$ is decreased to the minimal $\sigma'$ for which a $\sigma$-supermajority of $P$ voted for a value $\leq \sigma'$.
    (3) 
    If neither condition holds, $\sigma$ remains unchanged.

For example, changing $\sigma$ from $\frac{2}{3}$ to $\frac{3}{4}$ requires the approval of a $\frac{3}{4}$-supermajority.  Changing to unanimity requires approval by all members of $P$.

The $h$-rule is self-referential by design: raising the threshold to $\sigma'$ requires a $\sigma'$-supermajority supporting it (preventing a minority from imposing stricter requirements), while lowering requires the \emph{current} $\sigma$-supermajority to agree (preventing a minority from weakening the constitution).  Abramowitz et al.~\cite{abramowitz2021beginning} show that the $h$-rule is the unique rule satisfying a natural set of axioms for this self-referential setting.


\mypara{(3) Amending $\Delta$}
Each agent $i \in P$ states its preferred $\Delta_i \in \mathbb{R}^+$.  Since $\Delta$ is a one-dimensional, single-peaked domain with potential Sybil presence, we use the \emph{Suppress Outer-$f$} parameter update rule of Shahaf et al.~\cite{shahaf2019sybil}:\footnote{Equivalently, we can use the Reality-Enforcing rule of Meir et al.~\cite{meir2022sybil}, which is defined differently but is mathematically equivalent.  In the original reference~\cite{shahaf2019sybil} the rule refers to `outer-$\sigma$'; we use `outer-$f$' here to avoid confusion with our use of $\sigma$.}
\begin{itemize}
    \item If the median vote $> \Delta$: discard the maximal $f$ votes, compute the median $\Delta'$ of the remaining votes, and update to $\Delta'$ if $\Delta' > \Delta$.
    \item Symmetrically, if the median vote $< \Delta$: discard the minimal $f$ votes, compute the median $\Delta'$ of the remaining votes, and update to $\Delta'$ if $\Delta' < \Delta$.
    \item Otherwise, $\Delta$ remains unchanged.
\end{itemize}
The key properties: if all Sybil agents wish to increase $\Delta$ (to slow down the protocol) but the correct agents do not, $\Delta$ will not increase; symmetrically for decreasing.  If all correct agents unanimously wish to update $\Delta$ to some $\Delta'$, the update succeeds regardless of adversarial votes, provided $f < \frac{n}{3}$.


\begin{tcolorbox}[colback=gray!5!white,colframe=black!75!black,top=2pt,bottom=2pt] 
A constitution $(P,\sigma,\Delta)$ is amended democratically to $(P',\sigma',\Delta')$ via a periodic cycle, using the prevailing $(P,\sigma,\Delta)$ as status quo:
\begin{enumerate}
    \item $P \to P'$: per-member $\sigma$-\textbf{supermajority} among $P$, plus \textbf{consent of all new members} in $P' \setminus P$~\cite{shahaf2019sybil}.
    \item $\sigma \to \sigma'$: the \textbf{$h$-rule} for constitutional amendment~\cite{abramowitz2021beginning}.
    \item $\Delta \to \Delta'$: the \textbf{Suppress Outer-$f$} parameter update rule~\cite{shahaf2019sybil}.
\end{enumerate}
\end{tcolorbox}

%% file: formal_definitions.tex
\section{Formal Definitions}
\label{section:formal}

\subsection{Blocklace Preliminaries}
\label{section:prelim}

We use $a\ne b \in X$ as a shorthand for $a\in X \wedge b\in X\wedge a\ne b$.

A \emph{$p$-block} $b$, with $p\in \Pi$, is a triple $b=(h,x,H)$ where $h$ is the hash of the pair $(x,H)$ signed by $p$, referred to as the \emph{identifier} of $b$, $x$ is an arbitrary \emph{payload}, which may be \emph{empty}, $x=\bot$, in which case we refer to $b$ as an \emph{empty block}, and $H$ a finite set of block identifiers, the blocks of which are \emph{pointed to} by $b$.
A block with no pointers, $H=\emptyset$, is called \emph{initial}, in which case it can also be written as a pair $b=(h,x)$.
An encoding of a block via a sequence of bits in an agreed-upon form is referred to as a \emph{well-formed block}.
We assume the hash function to be collision-free whp and cryptographic, so that blocks cannot form cycles and we may identify a block $b$ with its identifier $h$. Furthermore, the method of signature allows the public key $p$ to be recovered from a signature by $p$.  

A set of blocks $X$ is referred to as a \emph{$\sigma$-supermajority given a set of agents $P$} if there is a $\sigma$-supermajority $Q\subseteq P$ such that for every $q\in Q$, $X$ includes at least one $q$-block.

A block $b$ \emph{observes} itself, any block $b'$ pointed to  by $b$, and any block observed by $b'$.  The `observes' relation induces a partial order on any set of blocks.
Two blocks that do not observe each other are \emph{conflicting}, and  
if by the same agent they are \emph{equivocating}. 
A block $b$ \emph{approves} a block $b'$ if $b$ observes $b'$ and does not observe any block equivocating with $b'$, and a set of blocks $B$ \emph{approves} $b'$ if every $b\in B$ approves $b'$.

The \emph{closure} of a block $b$, denoted $[b]$, is the set of blocks observed by $b$. 
For a set of blocks $B$, its \emph{closure} $[B]$ is defined by $[B] := \bigcup_{b\in B} [b]$, which is also the set of blocks \emph{observed by} $B$.
A set of blocks $B$ is \emph{closed} if $[b]\subseteq B$ for every $b\in B$, equivalently if $B=[B]$. 
A \emph{blocklace} is a closed set of blocks.  

To help avoid confusion, we use $B$ to denote a closed set of blocks (a blocklace) and $D$ or $X$ to denote any set of blocks, not necessarily closed.
Also, we refer to a set of blocks by agents in $P\subset\Pi$ as \emph{$P$-blocks}, and we let \emph{the $P$-blocks in $B$} denote the set of all blocks in $B$ by agents in $P$.

Given a set of blocks $D$, a pointer to a block $b$ is \emph{dangling} in $D$ if $b\notin D$.  Thus, a blocklace is a set of blocks with no dangling pointers. A block $b\in D$ is a \emph{tip} of $D$ if no other block $b'\ne b\in D$ observes $b$.

The \emph{depth} of a block $b$,  $\textit{depth}(b)$, is $0$ if $b$ is initial else $\textit{depth}(b)=\textit{depth}(b')+1$ where $b'$ is a maximal-depth block pointed to by $b$.   The \emph{depth}  of a set of blocks $B$ is defined by $\textit{depth}(B) := \max_{b\in B} \textit{depth}(b)$.  In the context of blocks, we use \emph{most-recent} as synonym for deepest.
A \emph{round} $r\ge 0$ in  $B$ is the set of all blocks $b\in B$ with $\textit{depth}(b)=r$, 
and the \emph{$r$-prefix of $B$}, $B_r$, consists of all blocks $b\in B$ with  $\textit{depth}(b)\le r$, equivalently all rounds in $B$ up to and including $r$.  Note that blocks of the same round do not observe each other. 

\subsection{Protocol concepts}
\label{section:concepts}

\mypara{Overview}
We begin by disregarding \emph{equivocations}, namely, two blocks by the same agent that do not observe each other.
We say that a block $b$ \emph{approves} a block $b'$ if $b$ observes $b'$ and does not observe any block equivocating with $b'$. We use this notion to define endorsements and ratification.

A second-round block $b$ of a wave may approve multiple first-round blocks in its wave, but {endorses} (towards finality) at most one of them, depending on the mode. The mode is determined according the
{quiescence} property of the wave $W$ that precedes $b$ in $[b]$, which we shall define later.
If $W$ is quiescent and $b$ approves exactly one round-one block in its wave, $b'$, then $b$ \emph{endorses} $b'$.
If $W$ is non-quiescent and $b$ approves a formal leader's round-one block, $b'$, in its wave, then $b$ \emph{endorses} $b'$.
In case neither condition is satisfied, $b$ does not endorse any block.

A first-round block $b$ is \emph{ratified} by a third-round block approving a $\sigma$-supermajority in $P$ of second-round blocks that endorse $b$. And a first-round block $b$ is \emph{final} once there is a  $\sigma$-supermajority in $P$ of third-round blocks that ratify $b$.

Figure \ref{fig:wave} illustrates a ``good case'' wave in which the leader block is finalised.
Observe that a wave can have at most one ratified or final block.
This is because a second-round block endorses at most one first-round block, and since correct agents do not equivocate, there cannot be  two $\sigma$-supermajorities in the same round endorsing different blocks.

For a given $P\subset \Pi$, the function \emph{leader$_P$} maps integers to $P$.  For each first round of a wave it assigns a \emph{formal leader} $p\in P$, where a first-round $p$-block $b$ is a \emph{leader block} if $p=\textit{leader}_P(\textit{depth}(b))$. We assume the leader function is  \emph{fair}, in that for every wave and every $p\in P$ there is a subsequent wave for which $p$ is the formal leader of its first round.

\begin{definition}[Endorse]
Let $b'$ be a second round block that approves a first round block $b$ in a wave $W$ in the blocklace $B=[b']$. Let $(P,\sigma)$ be the prevailing constitution.

Then $b'$ \temph{endorses} $b$ if the most-recent wave $W'  \subset B$ that precedes $b$ is:
\begin{enumerate}
    \item \textbf{quiescent} and 
        $b$ is the only first-round block approved by $b'$, or 
    \item \textbf{non-quiescent} and $b$ is a $\textit{leader}_P$ block approved by $b'$. 
\end{enumerate}
A set of second-round blocks $X\subset W$ \emph{endorses} $b$ if every $b'\in X$ endorses $b$.
\label{definition:endorse}
\end{definition}
Note that endorsement is an invariant block property, as it depends only on $[b']$.
Next, we specify the finality rule. A first-round block $b\in W$ is final if the third round of $W$ has a $\sigma$-supermajority that approves a second-round $\sigma$-supermajority that endorses $b$. Formally:

\begin{definition}[Ratified, Final, Ordered, Quiescent]\label{definition:final} 
Assume a blocklace $B$ and a wave $W\subseteq B$.
A first-round block $b\in W$ with prevailing constitution $(P,\sigma)$
is:
\begin{enumerate}
    \item \temph{ratified by} a third-round block $b'\in W$ if $b'$ approves a  $\sigma$-supermajority in $P$ that endorses $b$, in which case $b$ is also \temph{ratified in} $W$  and in $B$; and
    \item \temph{final} in $W$ and in $B$  if 
    the third-round of $W$ includes a  $\sigma$-supermajority in $P$, referred to as a \temph{finalising $\sigma$-supermajority}, in which each block ratifies $b$.
\end{enumerate} 
The wave $W$ is: 
\begin{enumerate}
    \item \temph{finalising} if it includes a block $b$ that is final in $W$; and 
    \item \temph{quiescent} if, in addition, all blocks in 
    $W$ except perhaps $b$ are empty, and no block in $B$ conflicts with $b$.
\end{enumerate}
In addition, the $0^{th}$-wave is \temph{quiescent} and its constitutional genesis block \temph{final}.
\end{definition}
Note that since only first-round blocks are endorsed, they are the only ones that can be ratified or final. Notice our use of the term ``finalised'' only applies to such leader blocks; the finalisation of leader blocks induces an order on all the blocks they approve, which are then said to be \emph{ordered}.
We observe that the block properties defined above are \emph{monotonic}:
\begin{observation}[Monotonicity]
If a block is ratified or final
in some blocklace $B$, it is also ratified or final,
respectively, in any blocklace $B'\supset B$.
\label{observation:monotonic-defs}
\end{observation}
\opodisCut{
Note, however, that quiescence is not monotonic, because conflicting and non-empty blocks may be added to a round. 
The next observation asserts that a wave's final block is unique.
}

To formalize the round progress rules, we define 
a round $r$ as \emph{advanced} in a blocklace $B$ if $B$ includes 
enough blocks to allow round $r+1$ blocks to be added to the blocklace safely:

\begin{definition}[Advanced Round]\label{definition:advanced-complete} 
Assume a constitutional blocklace $B$ with prevailing constitution $(P,\sigma)$.
A round $r$ in a wave $W\subseteq B$ is \temph{advanced} if it is a:
\begin{enumerate}
    \item first round that either  (\ia) includes a  $\sigma$-supermajority of blocks among $P$, or (\ib) includes a formal leader block or  (\ic) is preceded by a quiescent wave in $B$ and includes at least one block, or a
    \item second or third round that includes blocks by a $\sigma$-supermajority among $P$.
\end{enumerate}
Round $0$ is trivially advanced.
\end{definition}

Agents incorporate blocks in the blocklace only if they are valid, namely could have been produced by a correct agent during the run of the protocol. Formally:
\begin{definition}[Valid Block]\label{definition:valid-block}
A  block $b$ is \emph{valid} if it is well-formed and round $\textit{depth}(b)-1$ is advanced in $[b]$.
\end{definition}

\mypara{Dissemination} The protocol uses two types of \emph{dissemination-inducing blocks}:  \nack and  \inform;
other blocks are referred to as \emph{ordinary}.
Dissemination-inducing blocks are not incorporated in the blocklace of their sender or recipient. Hence, they contribute only to the liveness of the protocol and do not affect its safety, which is derived from structural properties of the blocklace. 
%
Unlike Cordial Miners, which incorporates a blocklace-based dissemination protocol with $O(n^2)$ amortized communication complexity in the good case, Constitutional Consensus achieves for this model $O(n)$  amortized communication complexity in the good case during high-throughput.


\opodisCut{
Protocols such as TCP achieve the reliable transfer of a sequence of bytes from the sender to the recipient by employing \textsc{ack} and \nack  messages for transmission control.
The purpose of an \textsc{ack} message is to acknowledge the receipt of a datagram, and of a \nack message to report an out-of-order datagram to its sender, e.g., in the case of receiving the $(k+1)^{th}$-datagram before receiving the $k^{th}$-datagram.

In the context of the blocklace, an `out-of-order' block is a block received before some block it points to, and the purpose of a \nack-block is indeed to report of such an out-of-order block to its creator (which is not necessarily its sender).  
}

%

\nack blocks are used to achieve reliable broadcast of ordinary blocks. They  are needed in order to rectify partial dissemination by Byzantine agents. For example, consider a Byzantine agent that sends a block $b$ to correct agent $p$ but not to another correct agent $q$, and $p$ adds $b$ into its blocklace and sends a block $b'$ that depends on $b$. Then when $q$ receives $b'$, it cannot add it to its blocklace because its predecessor $b$ is missing. To this end, $q$ sends a \nack block to $p$
specifying the blocks it is missing. 

Note that out-of-order blocks may naturally arise on a reliable network even during periods of synchrony and even when there are no Byzantine agents, because there is no bound on how \emph{quickly} messages can arrive --- in the example above, $b$'s sender might be correct and $b$ may still be underway to $q$. We therefore introduce a delay before sending a \nack. 
The result is that in the good case, when the network is synchronous and there is no partial-dissemination by Byzantine agents,  every correct agent sends its newly created blocks to all other agents and no \nack-blocks or other communication is needed.

The following definition relates to a setting in which an agent $p$ has a blocklace $B$ and it receives a new block $b$.  If $[b]\not\subseteq B\cup \{b\}$, then $p$ cannot incorporate $b$ into $B$ since some pointers in $b$ are dangling in $B$.   These are the tips of  $[b]\setminus (B\cup \{b\})$. 
\opodisCut{
Hence the following definition:
}

\begin{definition}[\nack-block]
Given a blocklace $B$, 
a \nack-\temph{block} by $p$ for a block $b=(h,x,H)$ such that $[b]\not\subseteq B\cup \{b\}$ is the $p$-block $b'=(h',x',H')$ where $x'=(\nack,h)$ and 
$H'\ne \emptyset$ contains the tips of  $[b] \setminus (B\cup \{b\})$. 
\end{definition}
Note that as defined above, $[b']\subseteq [b]$, and if $b$ has no pointers dangling in $B$ then a \nack-block for $b$ and $B$ is undefined.   If the \nack-block $b'$ is defined, then its pointers $H'$ are precisely the non-empty subset of $H$ that is dangling in $B$.

\opodisCut{
The reason for the asymmetry of having \nack-blocks but not \textsc{ack}-blocks is that any ordinary $p$-block functions as an \textsc{ack}-block, as it attests to all the blocks known to $p$ at the time of its creation.}
%
\nack blocks enable pull-based forwarding, where an agent who knows it misses a block asks for it. In addition, we make use of push-based forwarding
to facilitate leader progress using  \inform-blocks. 
\begin{definition}[\inform-Block]
Given a blocklace $B$, an \emph{ \inform-block} $(h,x,H)$ for round $r+1>1$ has payload $x= \inform$ and pointers $H$ to the blocks of depth $r$ in $B$.
\end{definition}

If an agent $p$ suspects that the leader of the next round $r+1$ does not proceed since it is missing some round $r$ blocks known to $p$, then $p$ sends the leader an \inform-block pointing to the $r$ blocks in its blocklace.   If the leader indeed misses any of these blocks, which may happen even after GST due to Byzantine partial dissemination,
it will respond to $p$ with a \nack-block, which will in turn cause $p$ to send to the leader the blocks it needs.

\remove{

\begin{definition}[Valid block]
    An ordinary block is \temph{valid} if it is well-formed and satisfies one of the following conditions:
    \begin{enumerate}
        \item it points to a $\sigma$-supermajority in round $\textit{depth}(b)-1$ in $[b]$, or
        \item it is a second-round block that points to a first-round leader block, or
        \item it is a second-round block and the predecessor wave in $[b]$ is quiescent. 
    \end{enumerate}
\end{definition}
}

\subsection{The Ordering Function $\tau$}
\label{appendix:tau}

\begin{definition}[$\tau$]\label{definition:tau}
	Assume a safe constitution $(P,\sigma)$ and any fixed topological sort function $\textit{xsort}(b,B,B')$ (exclude and sort) that
	takes a block $b$ and two blocklaces $B\supset B'$, and returns a sequence  that preserves the `observes' partial order and includes all the non-empty blocks in $B\setminus B'$ that are approved by $b$.  
    The function $\tau$ takes a block as input and returns a sequence of blocks in $[b]$ as output, as follows:
	$$
	\tau(b) :=
	\begin{cases}
        \tau(b') \cdot \textit{xsort}(b,[b],[b'])
        \text{\ \   if $b'$ is the most-recent block ratified in $[b]$}\\
		\textit{xsort}(b,[b]) \text{\ \ \ \ \   if no such $b'$ exists
       } \\

	\end{cases}
	$$
\end{definition}
When $\tau$ is called with a block $b$, it makes a recursive call with a block $b'$ ratified in $[b]$.
\opodisCut{
Note that $b'$ is not necessarily final in $[b]$, nor final in the closure of the final block with which $\tau$ was called initially. }
In the implementation of $\tau$, the results of recursive calls can be cached and used in subsequent calls, saving the need to recompute them.

%% file: pseudocode_walkthrough.tex
\begin{algorithm}[h]
\begin{tcolorbox}[colback=gray!5!white,colframe=white!75!white,top=2pt,bottom=2pt]
\vspace{3pt}
Let $B:=B^e$ herein and $r$ be the maximal advanced round of the current $B^e$.
\begin{description}\setlength{\itemsep}{2pt}
\item[Input:] Upon input a transaction $T$, $\emph{payload}\leftarrow \emph{payload}\cup\{T\}$

\item[Receive:] Upon receipt of a $q$-block $b$, $q\in P$\\
    if $b$ is a \nack-block for a $p$-block then judiciously send $[b]$ to $q$,\\
    else if $b$ an \inform-block and $[b]\not\subseteq B\cup \{b\}$\\
    \phantom{-------}  then send once a \nack-block for $b$ to $q$,\\
   else if $b$ is a valid ordinary block then $D \gets D \cup \{b\}$.

\item[Accept or Nack:] For every $q$-block $b\in D$, $q\in P$,\\
if $[b]\subseteq B\cup \{b\}$   then $B\gets B \cup \{b\}$, $D \gets D \setminus \{b\}$,\\ else
if a  $b$ was added to $D$ more than $\Delta$ time ago\\
    \phantom{-------}  then send once a \nack-block for $b$ to $q$.

\item[Issue:]  Issue once a round $r+1$ block if
$r+1$ is a second or third round or a first round that follows a wave that is:
\begin{enumerate}
    \item quiescent and $\emph{payload}\ne \bot$, or
    \item non-quiescent and (\ia) $p=\textit{leader}_P(r+1)$ or (\ib) $r$ has been advanced for $9\Delta$.
\end{enumerate}
\item[Issue Backlog:]
If $\emph{payload}\ne \bot$ and no round $r$ or round $r+1$
block has been issued by $p$, then issue once a round $r$ block.

\item[Inform Leader:] If $r$ is a third-round of a non-quiescent wave that has been advanced for $2\Delta$ then send once an \inform-block for round $r+1$ to $\textit{leader}_P(r+1)$.

\item[Output:]  If $B$ contains a final block $b$ of depth greater than the previous final block then output all the transactions in every block in $\tau(b)$ (Definition \ref{definition:tau}) that have not already been output.
\end{description}
\begin{itemize}
    \item \emph{issue a round $k$ block} means create a block $b$ with \textit{payload} and pointers to the tips of $B_{k-1}$, add $b$ to $B$,  send $b$ to every $q\ne p\in P$, and set $\emph{payload}\gets \bot$.

    \item \emph{judiciously send $[b]$ to $q$} means send every $b'\in [b]$ to $q$ unless (\ia) $p$ has already sent $b'$ to $q$, or (\ib)  $b'\in [b_q]$ for some $q$-block $b_q\in B\cup D$.
\end{itemize}
\end{tcolorbox}
\caption{Constitutional Consensus for a single epoch $e$ with prevailing constitution $(P,\sigma,\Delta)$.  Code for agent $p\in P$.}
\label{alg:consensus}
\end{algorithm}

\section{Pseudocode Walkthrough}\label{appendix:pseudocode-walkthrough}

{\bf  Receive:} valid incoming ordinary blocks are inserted to $D$.
Subsequently, the {\bf  Accept or Nack} is applied:
if any of their predecessors are missing, $p$ waits $\Delta$ to issue a \nack, and once a block in $D$ has no missing predecessors it is \emph{accepted} to $B$ and removed from $D$.

{\bf  Issue:}
The variable $r$ always holds the highest advanced round in $B$, and an agent typically issues blocks in round $r+1$.
This means that rounds can be skipped if the agent has been slow sending while the blocklace has advanced. This is useful when an agent that has been offline for some time catches up with the current blocklace before beginning to send blocks of its own. The only exception (to sending in round $r+1$) is
the {\bf  Issue Backlog} rule: In case $p$ has payload to send,   has not sent a round $r$ block, and is not supposed to send a round $r+1$ block, $p$  issues a round $r$ block to ensure its payload is disseminated.

If $r$ is a first or second round in its wave, then $p$ immediately (upon $r$ advancing) issues a round $r+1$ block. If $r$ is a third round, then block issuing (in the ensuing wave) depends on the mode. During low throughput (i.e., following a quiescent wave), $p$ issues a block if and only if it has payload to send.
During high-throughput, $p$ issues a block in case (\ia) it is the formal leader, or (\ib) it times-out on the leader.
Here, a long timeout is used ($9\Delta$), before which we inform the leader that the round has advanced through the
{\bf  Inform Leader} rule: Upon timing out -- with a short timeout of $2\Delta$ -- on the formal leader in the first round of a  high-throughput wave, $p$ sends an inform block to the formal leader in order to cause it to advance in case it is correct.

%% file: latency_complexity.tex
\section{Latency and Communication Complexity}\label{appendix:latency-complexity}

\mypara{Latency}
After GST and in the good case a leader block (formal or spontaneous) is finalised by its wave within $3\delta$, during both low- and high-throughput (See Figure \ref{fig:wave}).

\mypara{Communication complexity}
During low throughput, in the good case
the spontaneous leader block is sent to all, constant-size endorsement blocks are sent all-to-all, and the culprit is the third round, where blocks each with $O(n)$-pointers are sent to all-to-all.  In Morpheus~\cite{lewis2025morpheus}, this issue is alleviated by having second-round blocks be replaced by threshold signatures on the first block (in low-throughput mode): a $\sigma$-supermajority of these signatures can be amalgamated to form a single signature of constant length, and then third-round blocks can be replaced with threshold signatures on that single signature, giving communication complexity $O(n^2)$.

A similar idea can be applied here, by having hitherto-empty blocks carry a threshold signature instead.
Specifically, a second-round block, instead of being empty, should carry as payload a threshold signature of the pointer to the first-round leader block.  A third-round block that observes a supermajority of second-round blocks with threshold signatures, instead of being empty and pointing to all second-round blocks, would carry as payload an amalgamation of these threshold signatures, and point only to its own second-round block.
Thus, in the good case, all blocks of a low-throughput wave are of constant size,  resulting in $O(n^2)$ communication complexity.
\opodisCut{
To accommodate this change, the definition of a third-round $p$-block $b$ \emph{ratifying} a first-round leader block $b'$ has to include the case that $b$  points only to the second round $p$-block, but its payload is an amalgamation of threshold signatures of a  supermajority of second-round blocks that endorse $b'$.                  }
During high-throughput,  all blocks carry payloads.  If the payloads have at least $O(n)$ transactions each, then, in the good case,  the amortized communication complexity per transaction is $O(n)$.

%% file: single_epoch_delays_walkthrough.tex
\section{Single-Epoch Protocol: Delays}\label{appendix:single-epoch-walkthrough}

\mypara{Delays} The protocol makes judicious use of delays, to ensure both liveness and efficiency. Here is the rationale for the various protocol delays, all expressed as multiples of $\Delta$:

\mypara{$\Delta$ for issuing \nack-blocks}
In the good case we wish  no \nack-blocks to be sent at all. Agents wait $\Delta$ before complaining, so if a correct agent  sends a block $b$ to $p$ and $q$ at the same time, $q$ may receive $b$ up to $\Delta$ sooner than $q$ and send a block $b'$ that depends on $b$ to $p$; but $p$ will receive $b$ within $\Delta$ from receiving $b'$.
This delay is an optimization to avoid sending \nack-blocks in the good case; eliminating it does not hamper safety or liveness.
    
 \mypara{$2\Delta$ for informing the ensuing round's leader}
 In the good case whence all agents are correct after GST, whenever an agent receives a block, all other agents receive it within $\Delta$ time. Thus, once a third round is advanced at the blocklace of a correct non-leader agent $p$, $p$ expects to receive a leader block of the ensuing (first) round within $2\Delta$.  If this does not happen, this could be because (\ia) it is before GST, (\ib) the leader is incorrect, or (\ic) the leader has not seen yet that the previous round is advanced.  This last third case can happen after GST even if the leader is correct, due to Byzantine partial dissemination of third-round blocks.  Since $p$ knows that the previous third round is advanced, it sends an \inform-block to the leader, which will cause the leader to send a \nack-block requesting the blocks it is missing.  This $2\Delta$ delay is an optimization to avoid spurious  informs in the good case; eliminating it does not hamper safety or liveness.
    
 \mypara{$9\Delta$ for an unresponsive leader}   
    Following a non-quiescent wave after GST, if the leader $p_\ell$ is correct,  all correct agents should wait for $p_\ell$'s first round block before proceeding to the second round in order to allow that block to be finalized. But because $p_\ell$ may be faulty, they need to eventually time-out and stop waiting. To determine the time-out, we consider the worst-case latencies as follows: 
    Assume a third round $r$ is advanced at $p$ at time $t$. If $p$ does not receive a leader block by $t+2\Delta$, it  informs $p_\ell$, which issues a \nack-block, upon receipt of which $p$ sends $p_\ell$ any missing blocks. This exchange takes at most $3\Delta$,
    so by time $t+5\Delta$, $p_\ell$ issues a leader block $b$ for round $r+1$, which is received by $p$ by time $t+6\Delta$.  Upon receipt of $b$,  $p$ might be missing some blocks in $[b]$ and wait $\Delta$ before sending a \nack and getting them from $q$ by time $t+9\Delta$. If $p$ does not accept any leader-block by $t+9\Delta$, it concludes that either $p_\ell$ is faulty or GST has not arrived yet, and sends a first-round block of its own in order to allow the protocol to proceed to the next round. 

%% file: single-epoch-proof.tex
\section{Safety and Liveness of Single-Epoch Constitutional Consensus}\label{section:safety-liveness}


Here we prove the safety and liveness of the single-epoch Constitutional Consensus protocol presented in Algorithm \ref{alg:consensus}. In this case, the members of the founding constitution $C = (P,\sigma, \Delta)$ participate in the protocol for the entire run. The output sequence at every correct participant begins with $C$ and does not include additional constitutions.

\subsection{Safety}
\label{subsection:safety}

In this section, we prove the single-epoch protocol's safety.

\begin{theorem}\label{theorem-safety}
If the founding constitution is safe then single-epoch 
Constitutional Consensus is safe.        
\end{theorem}


The safety properties (Consistency and Validity) in Definition~\ref{definition:safety-liveness}  can be shown by proving the consistency of the output sequences of correct agents, formally defined as follows:
We say that two  sequences $x$ and $y$ are \emph{consistent} if one is a prefix of the other, namely there is a sequence $z$ such that $x\cdot z = y$ or  $x = y\cdot z$, with $\cdot$ denoting sequence concatenation.

\begin{observation}
If the output sequences of every two correct agents are consistent with each other, then single-epoch consensus ensures safety.
\label{observation:1-epoch-safety}
\end{observation}

\begin{proof}
Validity trivially holds because every agent participates in every epoch.  
Assuming the output sequences of correct agents are all consistent with each other, we take $T$ to be their supremum and get Consistency.   
\end{proof}

The next observation asserts that a wave's final block is unique.
\begin{observation}[Uniqueness]
A wave can have at most one ratified or final block.
\label{observation:unique-final}
\end{observation}
\begin{proof}
The observation follows from the fact that a second-round block endorses at most one first-round block in any blocklace it is included in, so to have two $\sigma$-supermajorities of the same round endorse two blocks, at least one correct agent must endorse two blocks in this round; to do so this correct agent must equivocate, a contradiction.
\end{proof}


To prove consistency, 
first, recall the assumption that in a blocklace $B$ created by the protocol 
with prevailing constitution $(P, \sigma)$, the $P$-blocks in $B$ include equivocations by at most $f = (2\sigma-1)n$ agents. We refer to blocklaces and waves satisfying this property as \emph{constitutionally-safe} wrt $(P, \sigma)$.


\begin{proposition}\label{observation:tau-no-equivocations}
If the prevailing constitution is safe then  $\tau(b)$ does not include equivocations.
\end{proposition}
\begin{proof}
Consider an equivocation $b_1, b_2 \in [b]$.  Consider the first call $\textit{xsort}(b',[b'],[b''])$ for which at least one of $b_1$ or $b_2$ is in $X:= [b']\setminus [b'']$, and the following two cases:
\begin{enumerate}
    \item Both $b_1,b_2 \in X$:  Then $b'$ approves neither hence neither will be in the output of $\textit{xsort}$ and hence nor in the output of $\tau$.
    \item Only one of them is in $X$, wlog $b_1\in X$, and since this is the first such call to $\textit{xsort}$ it must be that $b_2\in [b'']$.  Then  $b'$ observes both $b_1$ and $b_2$ and hence will approve neither, whereas $b''$ observes only $b_2$, therefore will approve it and therefore only $b_2$  will occur in the output of $\tau$, if at all (assuming there are no other equivocations with $b_2$).
\end{enumerate}
In either case, not both $b_1$ and $b_2$ will be in $\tau(b)$.
\end{proof}

For the following proofs we generalize $\tau$ to blocklaces, so that $\tau(B)$ is $\tau(b)$ if $b$ is the most-recent final block in $B$ if there is one, else it is the empty sequence $\Lambda$.

\begin{definition}[$\tau$ Safety]\label{definition:tau-safety}
A blocklace $B$ is \temph{$\tau$-safe} if for every wave $W\subset B$ with a block $b$ final in $W$ and any subsequent wave $W'\subset B$ with a block $b'$ ratified in $W'$, $b$ is ratified in $W\cap [b']$. 
\end{definition}

\begin{proposition}\label{proposition:tau-safety}
A constitutional blocklace with a safe constitution is $\tau$-safe. 
\end{proposition}
\begin{proof}
Let $B$ be a constitutional blocklace with a safe constitution $(P,\sigma)$, $W\subset B$ a wave with a block $b$ final in $W$, and $W'\subset B$ a subsequent wave with a block $b'$ ratified in $W'$.  We have to show that $b$ is ratified in $W\cap [b']$. 

Since $b'$ is ratified by some block it means that it is valid, since correct agents do not include invalid blocks in their blocklace, let alone endorse them, and a $\sigma$-supermajority includes (a majority of) correct agents.  Hence, being a first-round block of a wave, $b'$ must observe a $\sigma$-supermajority of the last round of the previous wave, let's call it the blue supermajority.
    
    
Then the blue supermajority and the finalising $\sigma$-supermajority 
    of $b$, let's call it the green supermajority, must have a correct agent in their intersection.  
    Namely, there is a correct agent $p\in P$ with a blue $p$-block that must observe a green $p$-block (which may actually be the same block if $W$ and $W'$ are consecutive waves), which in turn must observe a $\sigma$-supermajority of blocks that endorse $b$.
    Since $b'$ observes the blue $p$-block that observes the green $p$-block that ratifies $b$, then $b$ is ratified by $b'$.
\end{proof}

\begin{proposition}[$\tau$ Monotonicity]\label{proposition:tau-monotonic}
$\tau$ is monotonic wrt $\supset$ over constitutionally-safe blocklaces.
\end{proposition}
\begin{proof}
We have to show that if $B\subset B'$ for 
blocklaces $B, B'$ then $\tau(B)$ is a prefix of $\tau(B')$. 

By Observation~\ref{observation:monotonic-defs},
all the blocks final in $B$ are also final in $B'$. By Observation~\ref{observation:unique-final}, each wave in $B, B'$ includes at most one ratified or final block, hence the ratified and final blocks are totally ordered by their depth.
If $B, B'$ have the same most-recent final block, or if $B$ has no final block, then the proposition holds trivially.  
Assume $b\ne b'$ are the most-recent final blocks in $B, B'$, respectively, so necessarily $b'$ is more recent than $b$.
Let $b=b_1,\ldots,b_k=b'$, $k\ge 2$, be the sequence of all ratified blocks in $B'$ between and including $b$ and $b'$.  By Proposition \ref{proposition:tau-safety}  and the definition of $\tau$, $\tau(b')$ will eventually call recursively $\tau(b)$ and hence $\tau(b)$ is a prefix of $\tau(b')$, implying that  $\tau(B)$ is a prefix of $\tau(B')$.
\end{proof}

\begin{observation}\label{observation:union} 
If the union $B\cup B'$ of two blocklaces is constitutionally-safe then $\tau(B)$ and $\tau(B')$ are consistent.
\end{observation}
\begin{proof}
By monotonicity of $\tau$ over constitutionally-safe blocklaces (Proposition~\ref{proposition:tau-monotonic}), $\tau(B)$ and $\tau(B')$ are both prefixes of $\tau(B\cup B')$, which implies that one is a prefix of the other, namely they are consistent.
\end{proof}

\begin{proof}[Proof of Theorem \ref{theorem-safety}]

By assumption, the $P$-blocks in $B \cup B'$ include equivocations
by at most $f = (2\sigma-1)n$ agents, so this set is constitutionally-safe.
  Hence, by Observation \ref{observation:union}, the outputs of any two correct agents running Constitutional Consensus are always consistent. By Observation~\ref{observation:1-epoch-safety}, the protocol is safe.
\end{proof}

\subsection{Liveness}
\label{subsection:liveness}


Next, we discuss liveness. Having proved safety, we know that all output sequences of correct agents are subsequences of a common sequence. 
Let $T$ be the minimal such sequence. Then the set of elements in $T$ is the union of all the elements in output sequences of correct agents. 
Because all members of $P$ participate in all epochs, Epoch Liveness requires that all their output sequences include all the transactions in $T$, which are all the transactions output by correct agents. 
Constitution Amendment Liveness trivially holds in the single-epoch case. 
We conclude with the following observation:

\begin{observation}[Single-epoch Liveness]
\label{observation:1-epoch-liveness}
A single-epoch consensus protocol is live if and only if the output sequences of all correct participants include all the transactions input or output by correct participants. 
\end{observation}

\begin{theorem}\label{theorem-liveness}
Single-epoch Constitutional Consensus is live. 
\end{theorem}

We first prove that Constitutional Consensus disseminates all correct agents' blocks and the blocks they depend on.  

\begin{proposition}[Dissemination]\label{proposition:dissemination}
If a block $b$ is issued by a correct agent in Constitutional Consensus then $[b]$ is eventually received by all correct agents.
\end{proposition}
\begin{proof}
By way of contradiction assume that $p, q\in P$ are correct, $b_p$ is issued by $p$ (and received by $q$ since the network is reliable) and some block $b\in [b_p]$ is never received by $q$. 
Since by assumption $b$ (and possibly other blocks in $[b_p]$) have not been received by $q$, then after receiving $b_p$, $q$ issues a \nack-block $b_q$ for which $b\in [b_q]$. 
Upon receipt of $b_q$ by $p$, $p$ will send (once)  $[b_q]$, which includes $b$, to $q$.  Since the network is reliable, $q$ will eventually received $b$, a contradiction.
\end{proof}


The following observation follows directly from the definition of block validity:

\begin{observation}[Sequential Advance]
If a blocklace $B$ contains a round $r$  block, then every round $0< r'<r$ is  advanced in $B$. 
    \label{observation:sequential-advance}
\end{observation}

We next show that the protocol does not get ``stuck'', i.e., agents continue to produce blocks when they are needed for progress. Because Constitutional Consensus is quiescent, agents do not have to generate infinitely many blocks. But if some agent issues a  block in a round, that round eventually advances.

\begin{proposition}[Advance]
If a correct agent $p$ sends a block in some round $r$ in Constitutional Consensus, then $r$ eventually becomes advanced at every correct agent. 
    \label{proposition:advance}
\end{proposition}
\begin{proof}

Let $p_1$ be an agent that sends a block $b$ in round $r$.
Then when $p_1$ does so, $r-1$ is advanced in $p_1$'s blocklace. 

Assume by way of contradiction that $r$ does not advance at some correct agent $p_2$. By Observation~\ref{observation:sequential-advance}, no blocks of depth greater than $r$ are added to $p_2$'s blocklace.  By Proposition~\ref{proposition:dissemination}, this means that
no correct agent sends blocks in rounds $>r$.

Moreover, because $r$ does not advance at $p_2$, it must be the case that $p_2$ does not accept a supermajority of blocks in round $r$, which by Proposition~\ref{proposition:dissemination} implies that there is some correct agent $p_3$ that does not send a round $r$ block. 

By Proposition~\ref{proposition:dissemination}, $p_3$ receives all the blocks in $[b]$. Therefore, $p_3$ accepts all these blocks into its blocklace.  Thus, round $r-1$  is advanced in $p_3$'s blocklace. 

In case $r$ is a first round following a quiescent wave at $p_3$, $r$ advances at $p_3$ upon receipt of $b$, and $p_3$ issues a round $r+1$ block, a contradiction. 

Otherwise, $r$ is the maximum advanced round at $p_3$ and is not a first round following a quiescent wave at $p_3$. Therefore, $p_3$ issues a round $r$ block, a contradiction.
\end{proof}

We next prove liveness for finite runs.

\begin{proposition}[Finite implies quiescence]
Consider a run of Constitutional Consensus in which the blocklace $B$ of a correct agent $p$ is finite in a suffix of the run. Then $B$ is quiescent.

\label{proposition:finite-quiescence}
\end{proposition}
\begin{proof}
Assume by way of contradiction that $p$'s latest advanced round in the run, $r$, is not quiescent. 
Then, by the {\bf Issue} rule, $p$ eventually issues a block in round $r+1$ (at most $9\Delta$ time after $r$ is advanced). 

\end{proof}

\begin{proposition}[Block issuance]
If an agent $p$ has payload to send, then $p$ eventually issues a block with this payload.
\label{prop:issue}
\end{proposition}
\begin{proof}
Assume by contradiction that $p$ has a non-empty payload from some time when $p$'s most advanced round is $r_0$ onward, and $p$ never issues a block with this payload.
First, consider the case that $p$'s rounds do not advance in some suffix of the run. Then by Proposition~\ref{proposition:finite-quiescence},  $p$'s blocklace is quiescent. Hence, condition 1 of the {\bf Issue} rule is satisfied, and the payload is sent. A contradiction. Therefore, $p$'s advanced rounds  increase indefinitely. Consider a time when $p$'s highest advanced round $r > r_0 $ and the payload has not been sent. Then $p$  issues a block either in round $r+1$ (by the {\bf Issue} rule) or in round $r$ (by the {\bf Issue backlog} rule). A contradiction.
\end{proof}

\begin{proposition}[Transaction Liveness in finite runs]
Consider a run of Constitutional Consensus in which the blocklace $B$ of a correct agent is finite in a suffix of the run. Then all input transactions of correct agents are ordered in $B$.
    \label{proposition:quiescence}
\end{proposition}
\begin{proof}
By Proposition~\ref{proposition:dissemination}, $B$ includes all ordinary blocks sent by correct agents. By Proposition~\ref{proposition:finite-quiescence}, $B$ is quiescent.
This means that its last wave includes a first round block $b$ that is final in $B$, and all other blocks in the wave are empty.  Because $b$ is final, $p$ calls $\tau(b)$. Also by assumption of $B$ being quiescent, all non-empty ordinary blocks in $B$ are in $[b]$, and hence ordered by $\tau(b)$.  
Moreover, by Proposition \ref{prop:issue}, no agents have payloads to send. It follows that all input transactions have been sent in ordinary blocks and ordered.
\end{proof}


\begin{proposition}[Epoch Liveness in finite runs]
Consider a run of Constitutional Consensus in which the blocklace $B$ of a correct agent is finite in a suffix of the run. Then all transactions output by correct agents are ordered in $B$.
    \label{proposition:quiescence-output}
\end{proposition}
\begin{proof}
Consider a transaction $t$ output by a correct agent $p_1$ and another correct agent $p_2$ whose blocklace $B_2$ is finite in a suffix of the run. Because $t$ is at some point 
final in $p_1$'s blocklace, its  blocklace eventually includes a block $b$ with $t$ in its payload and a super-majority of blocks that observe $b$. This super-majority includes at least one $p_3$-block $b'$ by a correct agent $p_3$. By Proposition~\ref{proposition:dissemination}, $p_2$ $b'$ is included in $B_2$, and because $b \in [b]$, so is $b$.
Because $B_2$ is finite in a suffix of the run, by Proposition~\ref{proposition:finite-quiescence},
it is quiescent, which means that all non-empty blocks including $b$ are final in $B_2$, as needed. 
\end{proof}

Next, we show that after GST, all waves that have correct leaders in high throughput mode are finalising. Because quiescence is not monotonic, we consider here only waves $k$ for which the predecessor wave, $k-1$, is not quiescent at any correct agent at any time. 

\begin{proposition}[First round progress after GST]
Let $r$ be a first round with a correct leader, 
let $p$ be a correct agent at which round $r-1$ advances at time $t$>GST, 
and assume that the wave that ends in round $r-1$ is not quiescent in any correct agent's blocklace at any time. Then $p$ sends a block in round $r$ if and only if it is the round's leader. Moreover, $p$ endorses a round $r$ leader block in round $r+1$. 
\label{proposition:progress-first}
\end{proposition}

\begin{proof} 
Let $p_\ell$ be the correct leader of round $r$.
If $p=p_\ell$ then once $r-1$ is advanced,  $p$ immediately issues a leader block for round $r$, causing round $r$ to advance, and issues a block endorsing it in round $r+1$. 

Otherwise, $p$ waits for a block from $p_\ell$ until time $t+2\Delta$ and 
then sends $p_\ell$ an \inform-block, which is received by $t+3\Delta$, causing  $p_\ell$ to send $p$ a \nack-block and  get the blocks it is missing by time  $t+5\Delta$. 
Because round $r-1$ is advanced and its wave is not quiescent, $p_\ell$ sends a leader block, which is received by $p$ by time $t+6\Delta$. Some predecessors  of this block might be missing, causing $p$ to wait $\Delta$ and then send a \nack-block and get the missing blocks allowing it to accept the leader block by time $t+9\Delta$. 

When the leader block is received, the round advances and $p$'s round $r+1$ block endorses the leader block. 
\end{proof}

Below, we say that a round $r$ \emph{advances at time $t$} if there is some correct agent $p$ such that round $r$ is advanced in $p$'s blocklace at time $t$ and $r$ is not advanced at any correct agent before time $t$. 

\begin{proposition}[Progress after GST]
Let $r$ be a first round with a correct leader s.t.\ round $r-1$ advances  after GST and the wave that ends in round $r-1$ is not quiescent in any correct agent's blocklace at any time. Then a round $r$ leader block is finalized in its wave at all correct agents. 
\label{proposition:progress}
\end{proposition}

\begin{proof}
    By Proposition~\ref{proposition:progress-first}, the only correct agent that sends a round $r$ block is the round's leader and all correct agents endorse this block in round $r+1$. This means that it is impossible to create a valid round $r+1$ block that does not endorse the leader, as such a block would have to point to a super-majority of round $r$ blocks excluding the leader block.  
    Therefore, every valid round $r+2$ block ratifies the leader block, and the block is finalized in round $r+2$ as soon as it advances at any correct agent. 
\end{proof}

Next, we consider  runs with infinitely many quiescent waves.

\begin{proposition}[Infinite quiescent waves]
Consider a run of Constitutional Consensus in which infinitely many waves are quiescent in at least one correct agent's blocklace at some point in time. Then all the blocks of correct agents are ordered by all agents in this run. 
 \label{proposition:low-throughput}
\end{proposition}

\begin{proof}
Let $B$ be the (infinite) union of all blocklaces in the run. 
Let $b_1$ be a block of a correct agent in $B$. By Proposition~\ref{proposition:dissemination},  $b_1$ is eventually included in all correct agents' blocklaces.  Therefore, it is approved by all correct agents' blocks from some wave onward.
Because every quiescent wave is finalising, and finalization is monotonic,  there are infinitely many finalising waves in $B$.
Let $b_2$ be a final block that approves $b_1$ in some wave $k$. 
Once $b_2$ is final at some agent's blocklace, $b_1$ is ordered at that agent. 
We will show that $b_2$ is finalized at all agents. 

Observe that since $b_2$ is finalized in wave $k$, it is approved by a super-majority in wave $k$. Because every valid first round block approves a super-majority in the previous round and there is at least one correct agent at the intersection of any two super-majorities, every valid first round block in any round $k'>k$ observes $b_2$. 

Consider a finalising wave $k'>k$ in $B$ and let $b_3$ be the first-round block that is finalized in wave $k'$. 
Then $b_3$ is endorsed by a super-majority in the second round. 
Because this super-majority includes correct agents, all correct agents eventually accept $b_3$ (by Proposition~\ref{proposition:dissemination}). 
If $b_3$ is not a leader block, then to be validly endorsed, $[b_3]$ must be quiescent, and so upon receiving $b_3$, all the correct agents learn that $b_2$ is finalized.  

By the same token, if any non-leader first round block $b$ is endorsed by a valid block of any agent in the second round of wave $k+1$, then $[b]$ is quiescent. 
This, in turn, means that every third round block in the wave either ratifies the leader block or observes a block whose closure is quiescent. 

So when the third round of wave $k+1$ advances, every correct agent either observes a super-majority of blocks that ratify the leader block or observes a block whose closure is quiescent. In the former case the leader block is finalized, ordering all preceding ordinary blocks including $b_2$, and in the latter $b_2$ is already final in the block's closure.
\end{proof}

We are now ready to prove the liveness theorem.

\begin{proof}[Proof of Theorem \ref{theorem-liveness}]
By Observation~\ref{observation:1-epoch-liveness}, it suffices to show that every input or output transaction of a correct agent is eventually included in the output sequence of every correct agent. 
By Proposition \ref{prop:issue}, all input transactions in agents' payloads are eventually sent in (ordinary) blocks. Similarly,  every output transaction of a correct agent is included in some ordinary block. 
Let $b$ be a non-empty ordinary $p$-block included in some correct agent's blocklace in a run of Constitutional Consensus. 
If $p$ is correct, then by Proposition~\ref{proposition:dissemination}, every correct agent receives and incorporates $b$ in its local blocklace.   
Otherwise, $b$ is finalized in some blocklace, implying that it is approved by at least one $q$-block $b'$ by a correct agent $q$. Again, by  Proposition~\ref{proposition:dissemination}, every correct agent receives and incorporates every block in $[b']$, including $b$, in its local blocklace.

We consider three cases: 
First, assume the run is finite. By Propositions~\ref{proposition:quiescence} and~\ref{proposition:quiescence-output}, all non-empty ordinary blocks sent or output by correct agents in this run are ordered at all correct agents, and the theorem holds.
Second, consider a run in which infinitely many waves are quiescent in at least one correct agent’s blocklace at some point in time. By Proposition~\ref{proposition:low-throughput}, 
if $p$ is correct then $b$ is ordered, and otherwise $b'$ is ordered, implying that $b$ is ordered as well because $b\in [b']$. 

Finally, assume the run is infinite and contains only finitely many waves that are quiescent in at least one correct agent’s blocklace at some point in time. 
Let $k$ be a wave whose last round advances after GST and after which no wave is quiescent at any correct agent's blocklace. By fairness of leader selection, there are inifinitely many waves $k'>k$  with a correct leader.  By Proposition~\ref{proposition:progress}, a leader block is finalized at all correct agents in each of these waves, ordering all non-empty ordinary blocks from earlier rounds. 
\end{proof}

%% file: amendment_setting.tex
\section{The Amendment Setting}\label{appendix:amendment-setting}

We describe the general democratic operation.

\mypara{Unified proposing and voting}
Following Bulteau et al.~\cite{bulteau2021aggregation}, there is no proposal phase.  Whenever an agent decides to vote (afresh or as update), it issues a block with its timestamped and signed vote.
A vote is \emph{persistent}: if agent $i$ does not update its vote, the most recent one in the blocklace remains in effect.  The prevailing constitution thus always reflects the current preferences of all agents without requiring repeated re-voting.
As we show next, the decision spaces of $P$, $\sigma$, and $\Delta$ are all naturally embedded in corresponding metric spaces.

\mypara{Domain structure of the parameters}
The three constitutional parameters have distinct domain structures, which determine the appropriate aggregation rule for each:
\begin{itemize}
    \item \textbf{$P$ (population):}  For each potential participant $p$, each agent votes yes or no---effectively, a separate referendum on each member.  This is the simplest form of status-quo voting.

    \item \textbf{$\sigma$ (supermajority threshold):}  Agents state a preferred value in $[0.5, 1)$, a naturally single-peaked domain.  Additionally, $\sigma$ is \emph{self-referential}: it is the threshold used to amend the constitution, including $\sigma$ itself.  The amendment rule must therefore be consistent with its own output.

    \item \textbf{$\Delta$ (timeout):}  Agents state a preferred value in $\mathbb{R}^+$, again a single-peaked one-dimensional domain.
\end{itemize}

\remove{
\mypara{Constitutional amendment by consensus}
At a fixed periodic deadline (e.g., every Tuesday at midnight GMT),\footnote{Real-world realizations may wish to add response mechanisms for emergency situations.} participants use the consensus protocol to agree on the current set of votes.
At the $i^{th}$ epoch's deadline, each agent $p_j$ creates a transaction $T^j_i$ holding a set of signed votes including the latest vote it knows from each participant. The agent proposes this transaction as input to the single-epoch consensus protocol.

Once ordered, these transactions are not output to users but instead provided as input to the amendment protocol, run independently by each agent. Once the amendment protocol gets, for the first time, a $\sigma$-supermajority of such ordered transactions $T^j_i$ for an epoch $i$, it creates a set of votes that includes the latest vote from each participant included in any of the received sets. This set is used by the democratic decision process to determine the next constitution. Thanks to the consensus on vote sets and to the deterministic nature of the democratic decision process, all correct agents agree on the next constitution.
}

%% file: amendment_pseudocode.tex
\section{Constitutional Amendment Pseudocode}\label{appendix:amendment-pseudocode}

\begin{algorithm}[h]
\begin{tcolorbox}[colback=gray!5!white,colframe=white!75!white,top=2pt,bottom=2pt]
\vspace{3pt}
\begin{description}
\item[Amendment Finalisation by $p$:]
Upon issuing an ordinary block, if $p$'s amendment protocol has produced  a valid constitutional amendment decision $d$ that is not final in $B$ then issue the block with  $d$ instead of with \textit{payload}. 

\item[End Epoch by $p$:]
Upon  finalising in $B$ a constitutional block $b$ with constitutional amendment $d$, prevailing population $P$, and new population $P'$,  call $\tau(b)$,  create a $(\coronate,d)$ block, send it to $P\cup P'$, and cease participation in the epoch.


\item[Start Epoch by $q$:]  Let $d$ be a valid decision to amend the prevailing constitution, with new population $P'$.\\
Upon receipt by $q\in P'$ of a $\sigma$-supermajority among $P$ of $(\coronate,d)$ blocks, and  provided $q$ previously issued such a block if $q\in P$,
commence participation in the new epoch by $e \gets d$ as the new epoch constitutional genesis block, $B\gets B\cup \{d\}$, and retaining $D$ and \textit{payload}.

\item[Respond to Belated \nack by $q$:] Upon receipt of a \nack $q'$-block
     for a $q$-block $b$ of a past epoch $e$ for which $q'\in \Pi$ is a participant,  judiciously send $[b]$ to $q'$.
\end{description}
\end{tcolorbox}
\caption{Incorporating Constitutional Amendments in Constitutional Consensus.\\
Code for any participant $p$ in the prevailing constitution and for any $q\in \Pi$.}
\label{alg:ca}
\end{algorithm}

\mypara{Epoch transition walkthrough}
Algorithm~\ref{alg:ca} shows the additional logic supporting amendments in  Algorithm~\ref{alg:consensus}.
Every epoch commences with a constitutional genesis block that specifies its prevailing constitution.

{\bf Amendement finalisation:}
If, during the epoch, members of the prevailing population produce a valid decision $d$ to amend the prevailing constitution, they endeavour to finalise this decision by producing only constitutional blocks with $d$ as payload.

{\bf End Epoch:}
Once an agent finalises a constitutional block with $d$, it
produces a \coronate block with payload $(\coronate,d)$, sends it to both the old and new populations of $d$, and ceases participation in the  epoch.
Members of the old population that receive it and have yet to finalise $d$ can \nack any missing dependent blocks.
After ceasing to participate in the old epoch, an agent no longer adds blocks from that epoch to its blocklace. However, it must continue to respond to {\bf Belated \nack} requests from agents who are missing blocks in order to finalise the old epoch (due to Byzantine equivocation). An old epoch may be garbage-collected upon learning that all agents have coronated an ensuing  epoch.


{\bf Start Epoch:}
 A member of the new population that receives a $\sigma$-supermajority of \coronate blocks may commence the new epoch with $d$ as the constitutional genesis block. An agent that is a member of both the new and old populations  commences the new epoch after both
receiving a $\sigma$-supermajority of \coronate blocks
and sending one of its own.

The algorithm's output includes \emph{constitutional blocks} with constitutional amendment decisions as payload.
When $\tau$ orders a constitutional block with amendment $d$ for the first time, the protocol outputs the new constitution, suppressing subsequent copies.

%% file: amendment-proof.tex
\section{Safety and Liveness of Constitutional Amendments}\label{section:safety-constitutional}

\begin{theorem}[Constitutional Amendment Correctness]
    The Constitutional Consensus Algorithm (Algorithm
    \ref{alg:consensus}) with Constitutional Amendments (Algorithm \ref{alg:ca}) is correct. 
\end{theorem}

We need to prove that in every run of the protocol, 
there exists a sequence $T$ s.t.\ the two safety properties and two liveness properties of Definition~\ref{definition:safety-liveness} hold w.r.t.\ $T$. Note that the 
$i^{th}$ epoch $e_i$ is in fact $d_i$. 
We have proven that Algorithm \ref{alg:consensus} is correct in each epoch. Therefore, there is a sequence $T^i$ so that the output sequences produced by all participants in epoch $e_i$  are consistent and valid w.r.t. $T^i$. 

Let $T$ be the concatenation of all these sequences, $T := T^1 \cdot T^2 \cdots $. We next prove that the protocol satisfies safety and liveness properties w.r.t. $T$.
The theorem will follow.

\begin{lemma}[Constitutional Amendment Safety]
 Every run of the Constitutional Consensus Algorithm (Algorithm \ref{alg:consensus}) with Constitutional Amendments (Algorithm \ref{alg:ca}) satisfies safety w.r.t.\ $T$ defined above. 
\end{lemma}
\begin{proof}
Validity  immediately follows from the fact that agents participate only in epochs of which they are members. Consistency follows from the fact that agents 
participate in epochs sequentially, namely, they stop running the protocol in epoch $i$ before starting epoch $j$. 
 \end{proof}

\remove{
First we argue (Lemma) that the output of a non-final epoch $e$ is the same for all correct agents, namely $\tau(b)$ where $b$ is the block with which the constitutional amendment decision ends the epoch $e$.  The output of a final epoch is according to Theorem~\ref{theorem-safety}.

Let $T$ be a constitutional sequence of a run of an instance of Constitutional Consensus, and let $p\in \Pi$ be a correct agent.  If $p$ does not participate in any constitution in $T$ then the claim holds vacuously.  Let $d$ be a valid constitutional amendment decision in which $p$ is a member of the population of the new constitution.  If $d$ is initial, then $p$ simply creates a genesis block with $d$ and commences participation in the first epoch.
Otherwise, if $p$ receives a \coronate-block showing that the decision $d$ has been finalized, then
it creates a constitutional genesis block with $d$ and commences the epoch.  In either case its output would be as in the Lemma.

Therefore, the output of a correct agent would be the epochs it participates in.  Satisfying consistency and validity.
}


\begin{lemma}[Constitutional Amendment Liveness]
Every run of the Constitutional Consensus Algorithm (Algorithm \ref{alg:consensus}) with Constitutional Amendments (Algorithm \ref{alg:ca}) satisfies liveness w.r.t.\ $T$ defined above. 
\end{lemma}
\begin{proof}
Consider an epoch $e$ with prevailing constitution $(P,\sigma,\Delta)$, and the two liveness requirements. If $e$ is the last epoch then Epoch Liveness follows from the liveness of the single-epoch Algorithm~\ref{alg:consensus}, Theorem~\ref{theorem-liveness}.  Else assume $e$ is not the last epoch, and therefore it must have ended with a coronation by a $\sigma$-supermajority among $P$, all observing a wave finalizing a constitutional block $b$ with payload $d$.

Let $p\in P$ be a correct agent.  We argue that $p$ must eventually coronate $d$, and while doing so, call $\tau(b)$.  As all correct participants in the epoch $e$ end it by calling $\tau(b)$, all would output the same blocks for $e$.  If $p$ is among the $\sigma$-supermajority coronating $d$ we are done. 

Else there is at least one correct agent $q\in P$ that coronated $d$ and sent to $p$ a \coronate block $b'$ observing a wave $W$ that finalizes $b$.  Since the network is reliable, $p$ receives $b'$ but may be missing some blocks in $[b']$.  Then $p$ sends a $\nack(b')$-block to $q$, who responds (even after epoch $e$ has ended, thanks to the Respond to Belated \nack rule) by sending to $p$ the missing blocks in $W$ and their predecessors, upon receipt of which $p$ finalizes $b$, issues a $(\coronate,d)$ block, and outputs $\tau(b)$.

Considering Constitution Amendment Liveness, once a constitutional amendment $d$ is known, all correct participants follow the high-throughput protocol and produce only blocks with payload $d$.  Eventually a correct leader will finalize its block, following the same argument as in the proof of Theorem~\ref{theorem-liveness}, upon which all correct participants will eventually issue $(\coronate,d)$ blocks.
\end{proof}

%% file: grassroots.tex
\section{Constitutional Consensus is Grassroots}
\label{section:grassroots-is-grassroots}

We recall the formal definition of a grassroots protocol from reference~\cite{shapiro2025atomic},  instantiate it informally (bypassing the formalities of the definition in~\cite{shapiro2025atomic}) to the Constitutional Consensus protocol, and then argue that the protocol is grassroots.


\begin{definition}[Oblivious, Interactive, Grassroots]\label{definition:grassroots}
A  protocol $\calF$ is:
\begin{enumerate}
    \item \temph{oblivious} if for every $\emptyset \subset P \subset P' \subseteq \Pi$, 
    a run of the protocol over $P$ can proceed just the same in the presence of agents in $P'$. 
    \item  \temph{interactive} if for every $\emptyset \subset P \subset P' \subseteq \Pi$ and every  run of the protocol over $P$, the protocol can proceed from any configuration $c$ in $r$ to interact with members in $P'\setminus P$, namely produce a run that is not possible only over $P$.
    \item \temph{grassroots} if it is oblivious and interactive.
\end{enumerate}
\end{definition}

Note that the definition is stated in terms of any two populations  $\emptyset \subset P \subset P' \subseteq \Pi$.

Being oblivious, in the context of Constitutional Consensus, means that if the prevailing populations of the protocol during a run are always included in $P$, then the presence of additional agents hanging around, those in $P'\setminus P$, which do not participate in the protocol, should not interfere with the protocol or affect its behaviour.  Clearly, this is the case with Constitutional Consensus, as its prevailing populations may always choose not to amend the constitution to include members outside $P$.  Hence Constitutional Consensus satisfies the requirement of being oblivious.

Being interactive, in the context of Constitutional Consensus, means that if the prevailing populations of the protocol in some prefix of a run ending in a configuration $c$ are included in a set $P$, then the protocol has a computation starting from $c$ that interacts with members outside $P$ in a meaningful way, namely the protocol has a computation from $c$ that reaches a configuration $c'$, in which members of $P'\setminus P$ are included in the prevailing population, and thus $c'$ cannot be reached if the prevailing populations are all restricted to $P$.  Clearly, members of the prevailing population $P$, starting from configuration $c$, may amend the constitution to include in $P$ a new member $p'\in (P'\setminus P)$ and thus reach a  configuration $c'$ in which $p'\notin P$ is a member. Hence Constitutional Consensus satisfies the requirement of being interactive.  Together, these two properties imply that Constitutional Consensus is indeed grassroots. We thus conclude:
\begin{theorem}
Constitutional Consensus is grassroots.
\end{theorem}

Note that the two requirements cannot even be expressed in the context of the basic (no constitutional amendments) Constitutional Consensus protocol, as the population $P$ is given and fixed.  Also note that Constitutional Consensus being interactive critically-depends on the ability of the prevailing population $P$ to decide to add new members.

%% file: related_work_appendix.tex
\section{Related Work}\label{appendix:related-work}

Constitutional Consensus combines three properties: (\ia) democratic one-person-one-vote governance, (\ib) decentralised BFT consensus, and (\ic) self-governing constitutional amendment within the protocol. We survey prior work across five areas, corresponding to the five aspects of the protocol described in the introduction, and find that every existing system lacks at least one of these properties.

\subsection{Smart Contracts and Digital Social Contracts}

Deployed smart contract bytecode on Ethereum is immutable; amendments are achieved through proxy schemes that separate mutable state from swappable logic, controlled by admin keys, multisigs, or token-weighted governance~\cite{de2021smart}. Tezos~\cite{goodman2014tezos} embeds self-amendment at the protocol level, and the amendment mechanism is itself subject to amendment. In both cases, governance is plutocratic and amendments apply to code, not to a broader protocol constitution.

A digital social contract~\cite{cardelli2020digital} differs architecturally: it is a voluntary agreement among people, executed on the participants' own devices, and enforced by the participants themselves. Constitutional Consensus is designed for such contracts. Governance is democratic (one person, one vote) rather than plutocratic (one token, one vote), and amendments apply to the protocol constitution---including the participant set, supermajority threshold, and timing---not merely to code. These distinctions are summarised in Table~\ref{table:contracts}.

\begin{table}[!ht]
  \caption{Comparing Smart Contracts and Digital Social Contracts.}
  \label{table:contracts}
  \begin{center}
 \begin{tabular}{ | m{6.5em} | m{15em}| m{15em} | }
    \hline
     & \textbf{Smart Contract}~\cite{de2021smart}
 &  \textbf{Digital Social Contract}~\cite{cardelli2020digital}
\\
     \hline
    \hline
     \textbf{Among:} &  Accounts  &  People \\
    \hline
    \textbf{Executed by:} & Third-party validators & The people's smartphones \\
 \hline
  \textbf{Cost:} & Gas & None \\
 \hline
 \textbf{Consensus:} & Always required & Only when required by the contract \\
 \hline
 \textbf{Governance:} & Plutocratic (token-weighted voting) & Democratic (one person, one vote) \\
 \hline
 \textbf{Amended via:} & Proxy schemes and redeployment; bytecode is immutable & Constitutional amendments  \\
 \hline
 \textbf{Amendment authority:} & Token-weighted governance & Participants themselves via supermajority \\
 \hline
 \textbf{Applications:} & Decentralized Finance (DeFi)~\cite{schueffel2021defi,jensen2021introduction},\newline Decentralized Autonomous Organizations (DAOs)~\cite{ethereum:dao,ding2023survey},\newline  Non-Fungible Tokens (NFTs)~\cite{wang2021non} &
 Social graph and social networking~\cite{shapiro2023gsn},\newline grassroots currencies~\cite{shapiro2024gc,lewis2023grassroots},\newline democratic communities and their federation~\cite{shapiro2025GF}  \\
 \hline
  \end{tabular}
  \end{center}
\end{table}

\subsection{DAOs and Blockchain Governance}

Decentralised Autonomous Organisations~\cite{hassan2021decentralized} are the prevailing method for participant-governed digital organisations. Several blockchains can modify their own protocol rules through on-chain stakeholder voting: Tezos~\cite{goodman2014tezos} through a five-period procedure requiring an 80\% supermajority; Polkadot~\cite{wood2016polkadot} through forkless runtime upgrades via on-chain referenda; Cardano~\cite{cip1694} through a tripartite structure of Constitutional Committee, Delegated Representatives, and Stake Pool Operators; Cosmos~\cite{kwon2016cosmos} through its governance module; and the Internet Computer~\cite{camenisch2022ic} through its Network Nervous System. All are universally plutocratic:\footnote{Cardano is utilizing one-person-one-vote but only for its Constitutional Committee, which has essentially only veto power.} voting power is proportional to staked tokens. This is not accidental; stake-weighting provides built-in Sybil resistance, whereas democratic voting requires an external mechanism to establish unique personhood.

DAO frameworks built on existing blockchains are similarly plutocratic. Aragon~\cite{cuende2017aragon} defaults to ERC-20 token-weighted voting. Compound Governor~\cite{leshner2020compound} and OpenZeppelin Governor~\cite{openzeppelin2021governor} are built around ERC20Votes, tying governance power to token holdings. Snapshot~\cite{snapshot2020} offers configurable voting strategies including an equal-weight option, but is purely off-chain and admin-controlled. Optimism's Citizens' House~\cite{optimism2022governance} is a notable exception: it implements one-person-one-vote governance for retroactive public goods funding. However, the Citizens' House relies on Ethereum's proof-of-stake consensus, and the Optimism Foundation retains significant authority over proposal execution.

Schneider~\cite{schneider2022cryptoeconomics} argues that the economic logic embedded in blockchain systems systematically constrains democratic possibilities. Siddarth et al.~\cite{siddarth2020watches} identify the gap directly: ``the lack of a robust notion of personhood \ldots\ has led to the development of plutocracies.'' Talmon~\cite{talmon2023social} examines DAO governance from a computational social choice perspective. Buterin~\cite{buterin2021coinvoting} critiques coin voting as creating plutocratic governance vulnerable to vote buying and proposes proof-of-personhood alternatives. No DAO or self-amending blockchain achieves democratic governance with its own BFT consensus and self-governing constitutional amendment.

Consortium blockchain platforms such as Hyperledger Fabric~\cite{Androulaki2018Fabric} and R3~Corda~\cite{Brown2016Corda} implement organisational governance rather than democratic governance. In Fabric, configuration changes---including membership, block parameters, and governance policies---require signatures from a majority of organisation administrators, collected off-chain and submitted as configuration update transactions. Corda centralises governance further: a network operator controls certificate issuance and network parameters. While Fabric's hierarchical policy system bears structural resemblance to constitutional amendment (each policy specifies who can modify it), decision-making authority rests with designated administrators, not with individual participants exercising equal voting rights.

\subsection{Reconfiguration of Consensus Protocols}

Constitutional amendments generalise reconfiguration in two respects:
\begin{enumerate}
    \item \emph{Scope.} Reconfiguration refers to changing the set of active agents (servers, processes), whereas a constitution describes the entire consensus protocol, and in principle any aspect of it may be amended.
    \item \emph{Authority.} Even in self-governing systems like Raft~\cite{ongaro2014raft}, reconfiguration is a mechanism available to whoever holds the leader role. Constitutional amendments, by contrast, are decided by the participants themselves through a process specified by the constitution.
\end{enumerate}
In Constitutional Consensus, constitutional amendment preserves the protocol instance: the transaction history (the blocklace) carries over from one epoch to the next under the new constitution.  Starting a new instance, by contrast, would discard all prior state.  In this work, the amendable component of the constitution consists of $(P,\sigma,\Delta)$ (Definition \ref{definition:constitution}), parameters which affect the safety, liveness, and performance of almost any consensus protocol for eventual synchrony.

Lamport's Part-Time Parliament~\cite{lamport10.1145/279227.279229part-time} informally described reconfiguration as intrinsic: the legislators' decrees determine who constitutes the parliament. Lamport, Malkhi, and Zhou~\cite{lamport2010reconfiguring} later formalised the mechanism as Method~R$\alpha$, yet noted the idea ``still appears not to be well understood.'' No formal framework was given for \emph{who decides} on membership changes or how such decisions are governed.

The reconfiguration literature has focused on \emph{how} membership changes execute safely. Vertical Paxos~\cite{lamport2009vertical} introduced an auxiliary configuration master that determines the acceptor set for each configuration. DynaStore~\cite{aguilera2011dynamic} allows any client to propose reconfigurations but treats them as externally triggered. Spiegelman et al.~\cite{spiegelman2017dynamic} explicitly model an administrator invoking reconfiguration operations. SMART~\cite{lorch2006smart} handles crash-fault reconfiguration via an external coordinator. Raft~\cite{ongaro2014raft} treats configuration changes as special log entries committed through joint consensus of both old and new configurations. In the BFT setting, BFT-SMaRt~\cite{bessani2014bftsmart} provides the first practical BFT library with reconfiguration support, but relies on an external trusted third party for membership decisions. Duan and Zhang~\cite{duan2022foundations} showed that prior BFT reconfiguration approaches can compromise liveness, and introduced Dyno, the first formally analysed dynamic BFT protocol. Proof-of-Stake protocols achieve a form of self-reconfiguration, where validator sets change based on staking transactions~\cite{komatovic2025permissioned}, but governance is plutocratic.

ComChain~\cite{Vizier2020ComChain} and the subsequent SMaRtChain~\cite{Bessani2020FromByzantine} achieve BFT reconfiguration without a trusted third party: current participants run Byzantine consensus to decide on membership changes and employ a forgetting protocol that decouples permanent keys from per-view consensus keys to prevent removed nodes from creating forks. However, neither system provides a governance mechanism: the criteria for accepting new members are application-specific (e.g., proof-of-work, certification by an authority). Moreover, both are limited to membership changes and do not support amending protocol parameters or governance rules.

Across all of these, decision authority is either left unspecified, delegated to an external entity, determined by application-specific criteria, or determined plutocratically. The mechanism of reconfiguration is well-studied; the question of who decides---through democratic governance---is not.

\subsection{BFT Consensus Protocols}

PBFT~\cite{castro1999pbft} introduced the first practical algorithm for BFT state machine replication, tolerating up to $\lfloor(n-1)/3\rfloor$ Byzantine faults. HotStuff~\cite{yin2019hotstuff} achieved linear communication complexity with responsiveness and became the basis for the Diem blockchain. Tendermint~\cite{buchman2016tendermint} brought BFT consensus to blockchains and underlies the Cosmos network.

A second line of work replaces the leader-based approach with DAG-based protocols. DAG-Rider~\cite{keidar2021need} is the first asynchronous BFT Atomic Broadcast protocol achieving optimal resilience, amortised communication complexity, and time complexity simultaneously. Narwhal and Tusk~\cite{danezis2021narwhal} separate reliable transaction dissemination from ordering, enabling high throughput. Bullshark~\cite{giridharan2022bullshark} adds a practical low-latency synchronous fast path. Jolteon and Ditto~\cite{gelashvili2022jolteon} combine leader-based consensus with an asynchronous DAG fallback. Mysticeti~\cite{babel2023mysticeti} achieves the theoretical lower bound of three message delays for commit latency. Raikwar et al.~\cite{raikwar2024sokdag} provide a systematisation of knowledge on DAG-based consensus.

Constitutional Consensus builds on two protocols from this line of work: Cordial Miners~\cite{keidar2023cordial}, a DAG-based BFT Atomic Broadcast protocol, and Morpheus~\cite{lewis2025morpheus}, which adapts between quiescent and high-throughput modes. Both are used as building blocks; Constitutional Consensus adds constitutional amendment and democratic governance above them. None of the BFT protocols surveyed here addresses governance or self-amendment: they provide the consensus mechanism but leave the question of who participates and how rules change to external processes.

\subsection{Grassroots Protocols and Democratic Identity}

Democratic digital governance platforms achieve one-person-one-vote participation but rely on centralised infrastructure. Decidim~\cite{barandiaran2024decidim}, used by over 500 institutions, runs as a client-server web application. Loomio~\cite{loomio2024} provides equal-weight deliberation and voting, but all data resides on centralised servers. CONSUL Democracy~\cite{royo2020decidemadrid,consul2025}, serving approximately 250 governments, authenticates users via national ID but operates as a centralised web application. LiquidFeedback~\cite{behrens2018liquidfeedback} published a blockchain prototype with a swarm-behaviour-inspired consensus algorithm where node influence derives from the number of accredited persons trusting each node, and the roadmap envisions voting on application code. However, LiquidFeedback implements liquid democracy with transitive delegation rather than one-person-one-vote; its consensus is not characterised as BFT; and the self-amendment capability remained a roadmap item.

Proof-of-personhood (PoP) systems solve the Sybil resistance prerequisite for democratic governance. Ford~\cite{ford2020identity} argues that digital personhood---inalienable participation rights built on proof-of-personhood mechanisms---is a necessary foundation for digital democracy. PoPCoin~\cite{borge2017proof} uses pseudonym parties to enable equal-weight participation in ByzCoin consensus, achieving democratic identity with BFT consensus but not addressing self-governing amendment. Humanode~\cite{kavazi2023humanode} is the closest existing system to combining all three properties: it runs BFT consensus (BABE/GRANDPA on Substrate) with equal per-node weight, using biometric identity verification. However, its self-governance mechanism (Vortex DAO) operates as an off-chain simulation as of early 2026, with full authority transfer planned approximately four years after mainnet launch; Vortex is designed as a DAO overlay on the consensus protocol, not as constitutional amendment embedded within it. Idena~\cite{idena2018} achieves equal-weight consensus via flip-test identity verification, but its consensus is not formally characterised as BFT, and protocol upgrades are controlled by the development team. Encointer~\cite{brenzikofer2019encointer} provides proof of personhood through physical key-signing ceremonies but operates as a Polkadot parachain, inheriting its consensus.

Because Constitutional Consensus requires no external authority, global ledger, or staked capital, multiple independent instances can form, operate, and grow concurrently---the grassroots property~\cite{shapiro2023grassrootsBA,shapiro2025atomic}. This contrasts with Proof-of-Stake protocols like Ethereum~2.0~\cite{buterin2020combining}, whose security depends on the staked token having real-world value, creating a winner-take-all dynamic that precludes independent instances. Democratic platforms achieve one-person-one-vote but are centralised; PoP systems solve identity but remain building blocks. To the best of our knowledge, no existing system combines democratic one-person-one-vote governance, decentralised BFT consensus, and self-governing constitutional amendment into a single protocol.

%% file: mainConstitutionalConsensus_OPODIS.bbl

\begin{thebibliography}{78}


\ifx \showCODEN    \undefined \def \showCODEN     #1{\unskip}     \fi
\ifx \showDOI      \undefined \def \showDOI       #1{#1}\fi
\ifx \showISBNx    \undefined \def \showISBNx     #1{\unskip}     \fi
\ifx \showISBNxiii \undefined \def \showISBNxiii  #1{\unskip}     \fi
\ifx \showISSN     \undefined \def \showISSN      #1{\unskip}     \fi
\ifx \showLCCN     \undefined \def \showLCCN      #1{\unskip}     \fi
\ifx \shownote     \undefined \def \shownote      #1{#1}          \fi
\ifx \showarticletitle \undefined \def \showarticletitle #1{#1}   \fi
\ifx \showURL      \undefined \def \showURL       {\relax}        \fi
\providecommand\bibfield[2]{#2}
\providecommand\bibinfo[2]{#2}
\providecommand\natexlab[1]{#1}
\providecommand\showeprint[2][]{arXiv:#2}

\bibitem[Abramowitz et~al\mbox{.}(2021a)]%
        {abramowitz2021amend}
\bibfield{author}{\bibinfo{person}{Ben Abramowitz}, \bibinfo{person}{Ehud Shapiro}, {and} \bibinfo{person}{Nimrod Talmon}.} \bibinfo{year}{2021}\natexlab{a}.
\newblock \showarticletitle{How to Amend a Constitution? Model, Axioms, and Supermajority Rules}. In \bibinfo{booktitle}{\emph{Proceedings of AAMAS '21}}. \bibinfo{pages}{1443--1445}.
\newblock


\bibitem[Abramowitz et~al\mbox{.}(2021b)]%
        {abramowitz2021beginning}
\bibfield{author}{\bibinfo{person}{Ben Abramowitz}, \bibinfo{person}{Ehud Shapiro}, {and} \bibinfo{person}{Nimrod Talmon}.} \bibinfo{year}{2021}\natexlab{b}.
\newblock \showarticletitle{In the Beginning There Were $n$ Agents: Founding and Amending a Constitution}. In \bibinfo{booktitle}{\emph{Proceedings of ADT '21}}. \bibinfo{pages}{119--131}.
\newblock


\bibitem[Aguilera et~al\mbox{.}(2011)]%
        {aguilera2011dynamic}
\bibfield{author}{\bibinfo{person}{Marcos~K Aguilera}, \bibinfo{person}{Idit Keidar}, \bibinfo{person}{Dahlia Malkhi}, {and} \bibinfo{person}{Alexander Shraer}.} \bibinfo{year}{2011}\natexlab{}.
\newblock \showarticletitle{Dynamic atomic storage without consensus}.
\newblock \bibinfo{journal}{\emph{Journal of the ACM (JACM)}} \bibinfo{volume}{58}, \bibinfo{number}{2} (\bibinfo{year}{2011}), \bibinfo{pages}{1--32}.
\newblock


\bibitem[Androulaki et~al\mbox{.}(2018)]%
        {Androulaki2018Fabric}
\bibfield{author}{\bibinfo{person}{Elli Androulaki}, \bibinfo{person}{Artem Barger}, \bibinfo{person}{Vita Bortnikov}, \bibinfo{person}{Christian Cachin}, \bibinfo{person}{Konstantinos Christidis}, \bibinfo{person}{Angelo~De Caro}, \bibinfo{person}{David Enyeart}, \bibinfo{person}{Christopher Ferris}, \bibinfo{person}{Gennady Laventman}, \bibinfo{person}{Yacov Manevich}, \bibinfo{person}{Srinivasan Muralidharan}, \bibinfo{person}{Chet Murthy}, \bibinfo{person}{Binh Nguyen}, \bibinfo{person}{Manish Sethi}, \bibinfo{person}{Gari Singh}, \bibinfo{person}{Keith Smith}, \bibinfo{person}{Alessandro Sorniotti}, \bibinfo{person}{Chrysoula Stathakopoulou}, \bibinfo{person}{Marko Vukoli{\'c}}, \bibinfo{person}{Sharon~Weed Cocco}, {and} \bibinfo{person}{Jason Yellick}.} \bibinfo{year}{2018}\natexlab{}.
\newblock \showarticletitle{{Hyperledger Fabric}: A Distributed Operating System for Permissioned Blockchains}. In \bibinfo{booktitle}{\emph{Proceedings of the Thirteenth EuroSys Conference ({EuroSys})}}. \bibinfo{pages}{30:1--30:15}.
\newblock
\urldef\tempurl%
\url{https://doi.org/10.1145/3190508.3190538}
\showDOI{\tempurl}


\bibitem[Babel et~al\mbox{.}(2023)]%
        {babel2023mysticeti}
\bibfield{author}{\bibinfo{person}{Kushal Babel}, \bibinfo{person}{Andrey Chursin}, \bibinfo{person}{George Danezis}, \bibinfo{person}{Lefteris Kokoris-Kogias}, {and} \bibinfo{person}{Alberto Sonnino}.} \bibinfo{year}{2023}\natexlab{}.
\newblock \showarticletitle{Mysticeti: Low-Latency DAG Consensus with Fast Commit Path}.
\newblock \bibinfo{journal}{\emph{arXiv preprint arXiv:2310.14821}} (\bibinfo{year}{2023}).
\newblock


\bibitem[Barandiaran et~al\mbox{.}(2024)]%
        {barandiaran2024decidim}
\bibfield{author}{\bibinfo{person}{Xabier~E. Barandiaran}, \bibinfo{person}{Antonio Calleja-L\'{o}pez}, \bibinfo{person}{Arnau Monterde}, {and} \bibinfo{person}{Carol Romero}.} \bibinfo{year}{2024}\natexlab{}.
\newblock \bibinfo{booktitle}{\emph{Decidim, a Technopolitical Network for Participatory Democracy: Philosophy, Practice and Autonomy of a Collective Platform in the Age of Digital Intelligence}}.
\newblock \bibinfo{publisher}{Springer}.
\newblock
\urldef\tempurl%
\url{https://doi.org/10.1007/978-3-031-50784-7}
\showDOI{\tempurl}


\bibitem[Behrens et~al\mbox{.}(2018)]%
        {behrens2018liquidfeedback}
\bibfield{author}{\bibinfo{person}{Jan Behrens}, \bibinfo{person}{Axel Kistner}, \bibinfo{person}{Andreas Nitsche}, {and} \bibinfo{person}{Bj{\"o}rn Swierczek}.} \bibinfo{year}{2018}\natexlab{}.
\newblock \showarticletitle{The {LiquidFeedback} Blockchain}.
\newblock \bibinfo{journal}{\emph{The Liquid Democracy Journal}} \bibinfo{number}{6} (\bibinfo{year}{2018}), \bibinfo{pages}{18--29}.
\newblock
\showISSN{2198-9532}


\bibitem[Bessani et~al\mbox{.}(2020)]%
        {Bessani2020FromByzantine}
\bibfield{author}{\bibinfo{person}{Alysson Bessani}, \bibinfo{person}{Eduardo Ad{\'i}lio~Pelinson Alchieri}, \bibinfo{person}{Jo{\~a}o Sousa}, \bibinfo{person}{Andr{\'e} Oliveira}, {and} \bibinfo{person}{Fernando Pedone}.} \bibinfo{year}{2020}\natexlab{}.
\newblock \showarticletitle{From {B}yzantine Replication to Blockchain: Consensus is only the Beginning}. In \bibinfo{booktitle}{\emph{50th Annual {IEEE/IFIP} International Conference on Dependable Systems and Networks ({DSN})}}. \bibinfo{pages}{424--436}.
\newblock
\urldef\tempurl%
\url{https://doi.org/10.1109/DSN48063.2020.00057}
\showDOI{\tempurl}


\bibitem[Bessani et~al\mbox{.}(2014)]%
        {bessani2014bftsmart}
\bibfield{author}{\bibinfo{person}{Alysson Bessani}, \bibinfo{person}{Jo{\~a}o Sousa}, {and} \bibinfo{person}{Eduardo E.~P. Alchieri}.} \bibinfo{year}{2014}\natexlab{}.
\newblock \showarticletitle{State Machine Replication for the Masses with {BFT-SMART}}. In \bibinfo{booktitle}{\emph{2014 44th Annual IEEE/IFIP International Conference on Dependable Systems and Networks ({DSN})}}. \bibinfo{publisher}{IEEE}, \bibinfo{pages}{355--362}.
\newblock
\urldef\tempurl%
\url{https://doi.org/10.1109/DSN.2014.43}
\showDOI{\tempurl}


\bibitem[Borge et~al\mbox{.}(2017)]%
        {borge2017proof}
\bibfield{author}{\bibinfo{person}{Maria Borge}, \bibinfo{person}{Eleftherios Kokoris-Kogias}, \bibinfo{person}{Philipp Jovanovic}, \bibinfo{person}{Linus Gasser}, \bibinfo{person}{Nicolas Gailly}, {and} \bibinfo{person}{Bryan Ford}.} \bibinfo{year}{2017}\natexlab{}.
\newblock \showarticletitle{Proof-of-Personhood: Redemocratizing Permissionless Cryptocurrencies}. In \bibinfo{booktitle}{\emph{2017 {IEEE} European Symposium on Security and Privacy Workshops ({EuroS\&PW})}}. \bibinfo{publisher}{{IEEE}}, \bibinfo{pages}{23--26}.
\newblock
\urldef\tempurl%
\url{https://doi.org/10.1109/EuroSPW.2017.46}
\showDOI{\tempurl}


\bibitem[Brandt et~al\mbox{.}(2016)]%
        {brandt2016handbook}
\bibfield{author}{\bibinfo{person}{Felix Brandt}, \bibinfo{person}{Vincent Conitzer}, \bibinfo{person}{Ulle Endriss}, \bibinfo{person}{J{\'e}r{\^o}me Lang}, {and} \bibinfo{person}{Ariel~D Procaccia}.} \bibinfo{year}{2016}\natexlab{}.
\newblock \bibinfo{booktitle}{\emph{Handbook of computational social choice}}.
\newblock \bibinfo{publisher}{Cambridge University Press}.
\newblock


\bibitem[Brenzikofer(2019)]%
        {brenzikofer2019encointer}
\bibfield{author}{\bibinfo{person}{Alain Brenzikofer}.} \bibinfo{year}{2019}\natexlab{}.
\newblock \showarticletitle{Encointer -- Local Community Cryptocurrencies with Universal Basic Income}.
\newblock \bibinfo{journal}{\emph{CoRR}}  \bibinfo{volume}{abs/1912.12141} (\bibinfo{year}{2019}).
\newblock
\showeprint[arXiv]{1912.12141}
\urldef\tempurl%
\url{https://arxiv.org/abs/1912.12141}
\showURL{%
\tempurl}


\bibitem[Brown et~al\mbox{.}(2016)]%
        {Brown2016Corda}
\bibfield{author}{\bibinfo{person}{Richard~Gendal Brown}, \bibinfo{person}{James Carlyle}, \bibinfo{person}{Ian Grigg}, {and} \bibinfo{person}{Mike Hearn}.} \bibinfo{year}{2016}\natexlab{}.
\newblock \bibinfo{booktitle}{\emph{Corda: An Introduction}}.
\newblock \bibinfo{type}{{T}echnical {R}eport}. \bibinfo{institution}{R3}.
\newblock
\urldef\tempurl%
\url{https://docs.r3.com/en/pdf/corda-introductory-whitepaper.pdf}
\showURL{%
\tempurl}


\bibitem[Buchanan and Tullock(1962)]%
        {buchanan1965calculus}
\bibfield{author}{\bibinfo{person}{James~M. Buchanan} {and} \bibinfo{person}{Gordon Tullock}.} \bibinfo{year}{1962}\natexlab{}.
\newblock \bibinfo{booktitle}{\emph{The Calculus of Consent: Logical Foundations of Constitutional Democracy}}.
\newblock \bibinfo{publisher}{University of Michigan Press}.
\newblock


\bibitem[Buchman(2016)]%
        {buchman2016tendermint}
\bibfield{author}{\bibinfo{person}{Ethan Buchman}.} \bibinfo{year}{2016}\natexlab{}.
\newblock \emph{\bibinfo{title}{Tendermint: {Byzantine} Fault Tolerance in the Age of Blockchains}}.
\newblock Master of Applied Science Thesis. \bibinfo{school}{University of Guelph}.
\newblock
\urldef\tempurl%
\url{https://atrium.lib.uoguelph.ca/xmlui/handle/10214/9769}
\showURL{%
\tempurl}


\bibitem[Bulteau et~al\mbox{.}(2021)]%
        {bulteau2021aggregation}
\bibfield{author}{\bibinfo{person}{Laurent Bulteau}, \bibinfo{person}{Gal Shahaf}, \bibinfo{person}{Ehud Shapiro}, {and} \bibinfo{person}{Nimrod Talmon}.} \bibinfo{year}{2021}\natexlab{}.
\newblock \showarticletitle{Aggregation over Metric Spaces: Proposing and Voting in Elections, Budgeting, and Legislation}.
\newblock \bibinfo{journal}{\emph{Journal of Artificial Intelligence Research}}  \bibinfo{volume}{70} (\bibinfo{year}{2021}), \bibinfo{pages}{1413--1439}.
\newblock


\bibitem[Buterin(2021)]%
        {buterin2021coinvoting}
\bibfield{author}{\bibinfo{person}{Vitalik Buterin}.} \bibinfo{year}{2021}\natexlab{}.
\newblock \bibinfo{title}{Moving beyond coin voting governance}.
\newblock \bibinfo{howpublished}{Blog post}.
\newblock
\urldef\tempurl%
\url{https://vitalik.eth.limo/general/2021/08/16/voting3.html}
\showURL{%
\tempurl}


\bibitem[Buterin et~al\mbox{.}(2020)]%
        {buterin2020combining}
\bibfield{author}{\bibinfo{person}{Vitalik Buterin}, \bibinfo{person}{Diego Hernandez}, \bibinfo{person}{Thor Kamphefner}, \bibinfo{person}{Khiem Pham}, \bibinfo{person}{Zhi Qiao}, \bibinfo{person}{Danny Ryan}, \bibinfo{person}{Juhyeok Sin}, \bibinfo{person}{Ying Wang}, {and} \bibinfo{person}{Yan~X Zhang}.} \bibinfo{year}{2020}\natexlab{}.
\newblock \showarticletitle{Combining GHOST and casper}.
\newblock \bibinfo{journal}{\emph{arXiv preprint arXiv:2003.03052}} (\bibinfo{year}{2020}).
\newblock


\bibitem[Camenisch et~al\mbox{.}(2022)]%
        {camenisch2022ic}
\bibfield{author}{\bibinfo{person}{Jan Camenisch}, \bibinfo{person}{Andrea Cerulli}, \bibinfo{person}{David Derler}, \bibinfo{person}{Manu Drijvers}, \bibinfo{person}{Maria Dubovitskaya}, \bibinfo{person}{Jens Groth}, \bibinfo{person}{Timo Hanke}, \bibinfo{person}{Gregory Neven}, \bibinfo{person}{Yvonne-Anne Pignolet}, \bibinfo{person}{Victor Shoup}, \bibinfo{person}{Bj{\"o}rn Tackmann}, {and} \bibinfo{person}{Dominic Williams}.} \bibinfo{year}{2022}\natexlab{}.
\newblock \bibinfo{title}{The Internet Computer for Geeks}.
\newblock \bibinfo{howpublished}{Cryptology {ePrint} Archive, Paper 2022/087}.
\newblock
\urldef\tempurl%
\url{https://eprint.iacr.org/2022/087}
\showURL{%
\tempurl}


\bibitem[Cardelli et~al\mbox{.}(2020)]%
        {cardelli2020digital}
\bibfield{author}{\bibinfo{person}{Luca Cardelli}, \bibinfo{person}{Liav Orgad}, \bibinfo{person}{Gal Shahaf}, \bibinfo{person}{Ehud Shapiro}, {and} \bibinfo{person}{Nimrod Talmon}.} \bibinfo{year}{2020}\natexlab{}.
\newblock \showarticletitle{Digital social contracts: A foundation for an egalitarian and just digital society}. In \bibinfo{booktitle}{\emph{CEUR Proceedings of the First International Forum on Digital and Democracy}}, Vol.~\bibinfo{volume}{2781}. CEUR-WS, \bibinfo{pages}{51--60}.
\newblock


\bibitem[Castro and Liskov(1999)]%
        {castro1999pbft}
\bibfield{author}{\bibinfo{person}{Miguel Castro} {and} \bibinfo{person}{Barbara Liskov}.} \bibinfo{year}{1999}\natexlab{}.
\newblock \showarticletitle{Practical Byzantine Fault Tolerance}. In \bibinfo{booktitle}{\emph{3rd Symposium on Operating Systems Design and Implementation ({OSDI} 99)}}. \bibinfo{publisher}{{USENIX} Association}, \bibinfo{pages}{173--186}.
\newblock


\bibitem[{Consul Democracy Foundation}(2025)]%
        {consul2025}
\bibfield{author}{\bibinfo{person}{{Consul Democracy Foundation}}.} \bibinfo{year}{2025}\natexlab{}.
\newblock \bibinfo{title}{{CONSUL DEMOCRACY} --- The Most Complete Citizen Participation Tool for an Open, Transparent and Democratic Government}.
\newblock \bibinfo{howpublished}{\url{https://consuldemocracy.org/}}.
\newblock
\urldef\tempurl%
\url{https://consuldemocracy.org/}
\showURL{%
\tempurl}


\bibitem[Corduan et~al\mbox{.}(2023)]%
        {cip1694}
\bibfield{author}{\bibinfo{person}{Jared Corduan}, \bibinfo{person}{Andre Knispel}, \bibinfo{person}{Matthias Benkort}, \bibinfo{person}{Kevin Hammond}, \bibinfo{person}{Charles Hoskinson}, {and} \bibinfo{person}{Samuel Leathers}.} \bibinfo{year}{2023}\natexlab{}.
\newblock \bibinfo{title}{{CIP-1694}: A First Step Towards On-Chain Decentralized Governance}.
\newblock \bibinfo{howpublished}{Cardano Improvement Proposal}.
\newblock
\urldef\tempurl%
\url{https://github.com/cardano-foundation/CIPs/tree/master/CIP-1694}
\showURL{%
\tempurl}


\bibitem[Cuende and Izquierdo(2017)]%
        {cuende2017aragon}
\bibfield{author}{\bibinfo{person}{Luis Cuende} {and} \bibinfo{person}{Jorge Izquierdo}.} \bibinfo{year}{2017}\natexlab{}.
\newblock \bibinfo{title}{Aragon Network: A Decentralized Infrastructure for Value Exchange}.
\newblock \bibinfo{howpublished}{Whitepaper}.
\newblock
\urldef\tempurl%
\url{https://github.com/aragon/whitepaper}
\showURL{%
\tempurl}


\bibitem[Danezis et~al\mbox{.}(2021)]%
        {danezis2021narwhal}
\bibfield{author}{\bibinfo{person}{George Danezis}, \bibinfo{person}{Eleftherios~Kokoris Kogias}, \bibinfo{person}{Alberto Sonnino}, {and} \bibinfo{person}{Alexander Spiegelman}.} \bibinfo{year}{2021}\natexlab{}.
\newblock \showarticletitle{Narwhal and {T}usk: A {DAG}-based Mempool and Efficient {BFT} Consensus}.
\newblock \bibinfo{journal}{\emph{arXiv preprint arXiv:2105.11827}} (\bibinfo{year}{2021}).
\newblock


\bibitem[De~Filippi et~al\mbox{.}(2021)]%
        {de2021smart}
\bibfield{author}{\bibinfo{person}{Primavera De~Filippi}, \bibinfo{person}{Chris Wray}, {and} \bibinfo{person}{Giovanni Sileno}.} \bibinfo{year}{2021}\natexlab{}.
\newblock \showarticletitle{Smart contracts}.
\newblock \bibinfo{journal}{\emph{Internet Policy Review}} \bibinfo{volume}{10}, \bibinfo{number}{2} (\bibinfo{year}{2021}).
\newblock


\bibitem[Ding et~al\mbox{.}(2023)]%
        {ding2023survey}
\bibfield{author}{\bibinfo{person}{Qinxu Ding}, \bibinfo{person}{Daniel Liebau}, \bibinfo{person}{Zhiguo Wang}, {and} \bibinfo{person}{Weibiao Xu}.} \bibinfo{year}{2023}\natexlab{}.
\newblock \showarticletitle{A survey on decentralized autonomous organizations (DAOs) and their governance}.
\newblock \bibinfo{journal}{\emph{World Scientific Annual Review of Fintech}}  \bibinfo{volume}{1} (\bibinfo{year}{2023}), \bibinfo{pages}{2350001}.
\newblock


\bibitem[Duan and Zhang(2022)]%
        {duan2022foundations}
\bibfield{author}{\bibinfo{person}{Sisi Duan} {and} \bibinfo{person}{Haibin Zhang}.} \bibinfo{year}{2022}\natexlab{}.
\newblock \showarticletitle{Foundations of Dynamic {BFT}}. In \bibinfo{booktitle}{\emph{2022 {IEEE} Symposium on Security and Privacy ({SP})}}. \bibinfo{publisher}{{IEEE}}, \bibinfo{pages}{1317--1334}.
\newblock
\urldef\tempurl%
\url{https://doi.org/10.1109/SP46214.2022.9833787}
\showDOI{\tempurl}


\bibitem[Ethereum(ndao)]%
        {ethereum:dao}
\bibfield{author}{\bibinfo{person}{Ethereum}.} \bibinfo{year}{2021, https://ethereum.org/en/dao}\natexlab{}.
\newblock \bibinfo{title}{{Decentralized autonomous organizations (DAOs) | ethereum.org}}.
\newblock
\newblock
\urldef\tempurl%
\url{{https://ethereum.org/en/dao/}}
\showURL{%
\tempurl}


\bibitem[Ford(2020)]%
        {ford2020identity}
\bibfield{author}{\bibinfo{person}{Bryan Ford}.} \bibinfo{year}{2020}\natexlab{}.
\newblock \showarticletitle{Identity and Personhood in Digital Democracy: Evaluating Inclusion, Equality, Security, and Privacy in Pseudonym Parties and Other Proofs of Personhood}.
\newblock \bibinfo{journal}{\emph{CoRR}}  \bibinfo{volume}{abs/2011.02412} (\bibinfo{year}{2020}).
\newblock
\showeprint[arXiv]{2011.02412}
\urldef\tempurl%
\url{https://arxiv.org/abs/2011.02412}
\showURL{%
\tempurl}


\bibitem[Gelashvili et~al\mbox{.}(2022)]%
        {gelashvili2022jolteon}
\bibfield{author}{\bibinfo{person}{Rati Gelashvili}, \bibinfo{person}{Lefteris Kokoris-Kogias}, \bibinfo{person}{Alberto Sonnino}, \bibinfo{person}{Alexander Spiegelman}, {and} \bibinfo{person}{Zhuolun Xiang}.} \bibinfo{year}{2022}\natexlab{}.
\newblock \showarticletitle{Jolteon and Ditto: Network-Adaptive Efficient Consensus with Asynchronous Fallback}. In \bibinfo{booktitle}{\emph{Financial Cryptography and Data Security -- 26th International Conference, {FC} 2022}} \emph{(\bibinfo{series}{Lecture Notes in Computer Science}, Vol.~\bibinfo{volume}{13411})}. \bibinfo{publisher}{Springer}, \bibinfo{pages}{296--315}.
\newblock
\urldef\tempurl%
\url{https://doi.org/10.1007/978-3-031-18283-9\_14}
\showDOI{\tempurl}


\bibitem[Giridharan et~al\mbox{.}(2022)]%
        {giridharan2022bullshark}
\bibfield{author}{\bibinfo{person}{Neil Giridharan}, \bibinfo{person}{Lefteris Kokoris-Kogias}, \bibinfo{person}{Alberto Sonnino}, {and} \bibinfo{person}{Alexander Spiegelman}.} \bibinfo{year}{2022}\natexlab{}.
\newblock \showarticletitle{Bullshark: DAG BFT Protocols Made Practical}.
\newblock \bibinfo{journal}{\emph{arXiv preprint arXiv:2201.05677}} (\bibinfo{year}{2022}).
\newblock


\bibitem[Goodman(2014)]%
        {goodman2014tezos}
\bibfield{author}{\bibinfo{person}{LM Goodman}.} \bibinfo{year}{2014}\natexlab{}.
\newblock \showarticletitle{Tezos: A self-amending crypto-ledger position paper}.
\newblock \bibinfo{journal}{\emph{Aug}}  \bibinfo{volume}{3} (\bibinfo{year}{2014}), \bibinfo{pages}{2014}.
\newblock


\bibitem[Hassan and De~Filippi(2021)]%
        {hassan2021decentralized}
\bibfield{author}{\bibinfo{person}{Samer Hassan} {and} \bibinfo{person}{Primavera De~Filippi}.} \bibinfo{year}{2021}\natexlab{}.
\newblock \showarticletitle{Decentralized Autonomous Organization}.
\newblock \bibinfo{journal}{\emph{Internet Policy Review}} \bibinfo{volume}{10}, \bibinfo{number}{2} (\bibinfo{year}{2021}), \bibinfo{pages}{1--10}.
\newblock


\bibitem[{Idena Team}(2018)]%
        {idena2018}
\bibfield{author}{\bibinfo{person}{{Idena Team}}.} \bibinfo{year}{2018}\natexlab{}.
\newblock \bibinfo{title}{Idena: Proof-of-Person Blockchain}.
\newblock \bibinfo{howpublished}{Whitepaper}.
\newblock
\urldef\tempurl%
\url{https://docs.idena.io/docs/wp/summary/}
\showURL{%
\tempurl}


\bibitem[Jensen et~al\mbox{.}(2021)]%
        {jensen2021introduction}
\bibfield{author}{\bibinfo{person}{Johannes~Rude Jensen}, \bibinfo{person}{Victor von Wachter}, {and} \bibinfo{person}{Omri Ross}.} \bibinfo{year}{2021}\natexlab{}.
\newblock \showarticletitle{An introduction to decentralized finance (defi)}.
\newblock \bibinfo{journal}{\emph{Complex Systems Informatics and Modeling Quarterly}} \bibinfo{number}{26} (\bibinfo{year}{2021}), \bibinfo{pages}{46--54}.
\newblock


\bibitem[Kavazi et~al\mbox{.}(2023)]%
        {kavazi2023humanode}
\bibfield{author}{\bibinfo{person}{Dato Kavazi}, \bibinfo{person}{Victor Smirnov}, \bibinfo{person}{Sasha Shilina}, \bibinfo{person}{Jonathan Shomroni}, \bibinfo{person}{Rafael Contreras}, \bibinfo{person}{Hardik Gajera}, {and} \bibinfo{person}{Dmitry Lavrenov}.} \bibinfo{year}{2023}\natexlab{}.
\newblock \showarticletitle{Humanode: The First Crypto-Biometric Network: Decentralized Sybil Resistance through Biometric Identity Verification}. In \bibinfo{booktitle}{\emph{Proceedings of the 2023 4th Asia Service Sciences and Software Engineering Conference ({ASSE} 2023)}}. \bibinfo{publisher}{ACM}.
\newblock
\urldef\tempurl%
\url{https://doi.org/10.1145/3634814.3634843}
\showDOI{\tempurl}


\bibitem[Keidar et~al\mbox{.}(2021)]%
        {keidar2021need}
\bibfield{author}{\bibinfo{person}{Idit Keidar}, \bibinfo{person}{Eleftherios Kokoris-Kogias}, \bibinfo{person}{Oded Naor}, {and} \bibinfo{person}{Alexander Spiegelman}.} \bibinfo{year}{2021}\natexlab{}.
\newblock \showarticletitle{All you need is dag}. In \bibinfo{booktitle}{\emph{Proceedings of the 2021 ACM Symposium on Principles of Distributed Computing}}. \bibinfo{pages}{165--175}.
\newblock


\bibitem[Keidar et~al\mbox{.}(2023)]%
        {keidar2023cordial}
\bibfield{author}{\bibinfo{person}{Idit Keidar}, \bibinfo{person}{Oded Naor}, {and} \bibinfo{person}{Ehud Shapiro}.} \bibinfo{year}{2023}\natexlab{}.
\newblock \showarticletitle{Cordial Miners: A Family of Simple and Efficient Consensus Protocols for Every Eventuality}. In \bibinfo{booktitle}{\emph{37th International Symposium on Distributed Computing (DISC 2023)}} (Italy). \bibinfo{publisher}{LIPICS}.
\newblock


\bibitem[Komatovic et~al\mbox{.}(2025)]%
        {komatovic2025permissioned}
\bibfield{author}{\bibinfo{person}{Jovan Komatovic}, \bibinfo{person}{Andrew Lewis-Pye}, \bibinfo{person}{Joachim Neu}, \bibinfo{person}{Tim Roughgarden}, {and} \bibinfo{person}{Ertem~Nusret Tas}.} \bibinfo{year}{2025}\natexlab{}.
\newblock \showarticletitle{From Permissioned to Proof-of-Stake Consensus}. In \bibinfo{booktitle}{\emph{7th Conference on Advances in Financial Technologies ({AFT} 2025)}} \emph{(\bibinfo{series}{Leibniz International Proceedings in Informatics (LIPIcs)}, Vol.~\bibinfo{volume}{354})}. \bibinfo{publisher}{Schloss Dagstuhl -- Leibniz-Zentrum f{\"u}r Informatik}, \bibinfo{pages}{18:1--18:26}.
\newblock
\urldef\tempurl%
\url{https://doi.org/10.4230/LIPIcs.AFT.2025.18}
\showDOI{\tempurl}


\bibitem[Kwon and Buchman(2016)]%
        {kwon2016cosmos}
\bibfield{author}{\bibinfo{person}{Jae Kwon} {and} \bibinfo{person}{Ethan Buchman}.} \bibinfo{year}{2016}\natexlab{}.
\newblock \bibinfo{title}{Cosmos: A Network of Distributed Ledgers}.
\newblock \bibinfo{howpublished}{Whitepaper}.
\newblock
\urldef\tempurl%
\url{https://cosmos.network/whitepaper}
\showURL{%
\tempurl}


\bibitem[Lamport(1998)]%
        {lamport10.1145/279227.279229part-time}
\bibfield{author}{\bibinfo{person}{Leslie Lamport}.} \bibinfo{year}{1998}\natexlab{}.
\newblock \showarticletitle{The part-time parliament}.
\newblock \bibinfo{journal}{\emph{ACM Trans. Comput. Syst.}} \bibinfo{volume}{16}, \bibinfo{number}{2} (\bibinfo{date}{May} \bibinfo{year}{1998}), \bibinfo{pages}{133–169}.
\newblock
\showISSN{0734-2071}
\urldef\tempurl%
\url{https://doi.org/10.1145/279227.279229}
\showDOI{\tempurl}


\bibitem[Lamport(2001)]%
        {lamport2001paxos}
\bibfield{author}{\bibinfo{person}{Leslie Lamport}.} \bibinfo{year}{2001}\natexlab{}.
\newblock \showarticletitle{Paxos made simple}.
\newblock \bibinfo{journal}{\emph{ACM SIGACT News (Distributed Computing Column) 32, 4 (Whole Number 121, December 2001)}} (\bibinfo{year}{2001}), \bibinfo{pages}{51--58}.
\newblock


\bibitem[Lamport et~al\mbox{.}(2009)]%
        {lamport2009vertical}
\bibfield{author}{\bibinfo{person}{Leslie Lamport}, \bibinfo{person}{Dahlia Malkhi}, {and} \bibinfo{person}{Lidong Zhou}.} \bibinfo{year}{2009}\natexlab{}.
\newblock \showarticletitle{Vertical {P}axos and Primary-Backup Replication}. In \bibinfo{booktitle}{\emph{Proceedings of the 28th ACM Symposium on Principles of Distributed Computing (PODC)}}. \bibinfo{publisher}{ACM}, \bibinfo{pages}{312--313}.
\newblock


\bibitem[Lamport et~al\mbox{.}(2010)]%
        {lamport2010reconfiguring}
\bibfield{author}{\bibinfo{person}{Leslie Lamport}, \bibinfo{person}{Dahlia Malkhi}, {and} \bibinfo{person}{Lidong Zhou}.} \bibinfo{year}{2010}\natexlab{}.
\newblock \showarticletitle{Reconfiguring a State Machine}.
\newblock \bibinfo{journal}{\emph{SIGACT News}} \bibinfo{volume}{41}, \bibinfo{number}{1} (\bibinfo{year}{2010}), \bibinfo{pages}{63--73}.
\newblock


\bibitem[Lamport et~al\mbox{.}(1982)]%
        {lamport1982byzantine}
\bibfield{author}{\bibinfo{person}{Leslie Lamport}, \bibinfo{person}{Robert Shostak}, {and} \bibinfo{person}{Marshall Pease}.} \bibinfo{year}{1982}\natexlab{}.
\newblock \showarticletitle{The Byzantine generals problem}.
\newblock \bibinfo{journal}{\emph{ACM Transactions on Programming Languages and Systems}} \bibinfo{volume}{4}, \bibinfo{number}{3} (\bibinfo{year}{1982}), \bibinfo{pages}{382--401}.
\newblock


\bibitem[Leshner(2020)]%
        {leshner2020compound}
\bibfield{author}{\bibinfo{person}{Robert Leshner}.} \bibinfo{year}{2020}\natexlab{}.
\newblock \bibinfo{title}{Compound Governance}.
\newblock \bibinfo{howpublished}{Blog post, Compound Labs}.
\newblock
\urldef\tempurl%
\url{https://medium.com/compound-finance/compound-governance-5531f524cf68}
\showURL{%
\tempurl}


\bibitem[Lewis-Pye et~al\mbox{.}(2023)]%
        {lewis2023grassroots}
\bibfield{author}{\bibinfo{person}{Andrew Lewis-Pye}, \bibinfo{person}{Oded Naor}, {and} \bibinfo{person}{Ehud Shapiro}.} \bibinfo{year}{2023}\natexlab{}.
\newblock \showarticletitle{Grassroots Flash: A Payment System for Grassroots Cryptocurrencies}.
\newblock \bibinfo{journal}{\emph{arXiv preprint arXiv:2309.13191}} (\bibinfo{year}{2023}).
\newblock


\bibitem[Lewis-Pye and Shapiro(2025)]%
        {lewis2025morpheus}
\bibfield{author}{\bibinfo{person}{Andrew Lewis-Pye} {and} \bibinfo{person}{Ehud Shapiro}.} \bibinfo{year}{2025}\natexlab{}.
\newblock \showarticletitle{Morpheus Consensus: Excelling on trails and autobahns}.
\newblock \bibinfo{journal}{\emph{arXiv preprint arXiv:2502.08465}} (\bibinfo{year}{2025}).
\newblock


\bibitem[{Loomio Cooperative Limited}(2024)]%
        {loomio2024}
\bibfield{author}{\bibinfo{person}{{Loomio Cooperative Limited}}.} \bibinfo{year}{2024}\natexlab{}.
\newblock \bibinfo{title}{Loomio --- Make Decisions Together}.
\newblock \bibinfo{howpublished}{\url{https://www.loomio.com/}}.
\newblock
\urldef\tempurl%
\url{https://www.loomio.com/}
\showURL{%
\tempurl}


\bibitem[Lorch et~al\mbox{.}(2006)]%
        {lorch2006smart}
\bibfield{author}{\bibinfo{person}{Jacob~R. Lorch}, \bibinfo{person}{Atul Adya}, \bibinfo{person}{William~J. Bolosky}, \bibinfo{person}{Ronnie Chaiken}, \bibinfo{person}{John~R. Douceur}, {and} \bibinfo{person}{Jon Howell}.} \bibinfo{year}{2006}\natexlab{}.
\newblock \showarticletitle{The {SMART} Way to Migrate Replicated Stateful Services}. In \bibinfo{booktitle}{\emph{Proceedings of the 1st ACM SIGOPS/EuroSys European Conference on Computer Systems ({EuroSys} '06)}}. \bibinfo{publisher}{ACM}, \bibinfo{pages}{103--115}.
\newblock
\urldef\tempurl%
\url{https://doi.org/10.1145/1217935.1217946}
\showDOI{\tempurl}


\bibitem[Lynch(1996)]%
        {lynch1996distributed}
\bibfield{author}{\bibinfo{person}{Nancy~A. Lynch}.} \bibinfo{year}{1996}\natexlab{}.
\newblock \bibinfo{booktitle}{\emph{Distributed Algorithms}}.
\newblock \bibinfo{publisher}{Morgan Kaufmann}.
\newblock
\showISBNx{978-1-55860-348-6}


\bibitem[Marino and {Snapshot Labs}(2020)]%
        {snapshot2020}
\bibfield{author}{\bibinfo{person}{Fabien Marino} {and} \bibinfo{person}{{Snapshot Labs}}.} \bibinfo{year}{2020}\natexlab{}.
\newblock \bibinfo{title}{Snapshot: Off-chain Gasless Multi-governance Client}.
\newblock \bibinfo{howpublished}{Open-source software}.
\newblock
\urldef\tempurl%
\url{https://snapshot.org}
\showURL{%
\tempurl}


\bibitem[Meir et~al\mbox{.}(2022)]%
        {meir2022sybil}
\bibfield{author}{\bibinfo{person}{Reshef Meir}, \bibinfo{person}{Gal Shahaf}, \bibinfo{person}{Ehud Shapiro}, {and} \bibinfo{person}{Nimrod Talmon}.} \bibinfo{year}{2022}\natexlab{}.
\newblock \showarticletitle{Sybil-Resilient Social Choice with Partial Information}.
\newblock \bibinfo{journal}{\emph{Journal of Artificial Intelligence Research}}  \bibinfo{volume}{73} (\bibinfo{year}{2022}), \bibinfo{pages}{1517--1545}.
\newblock
\urldef\tempurl%
\url{https://doi.org/10.1613/jair.1.13390}
\showDOI{\tempurl}


\bibitem[Meir et~al\mbox{.}(2024)]%
        {meir2024safe}
\bibfield{author}{\bibinfo{person}{Reshef Meir}, \bibinfo{person}{Gal Shahaf}, \bibinfo{person}{Ehud Shapiro}, {and} \bibinfo{person}{Nimrod Talmon}.} \bibinfo{year}{2024}\natexlab{}.
\newblock \showarticletitle{Safe Voting: Resilience to Abstention and Sybils}.
\newblock \bibinfo{journal}{\emph{arXiv preprint arXiv:2001.05271}} (\bibinfo{year}{2024}).
\newblock


\bibitem[Ongaro and Ousterhout(2014)]%
        {ongaro2014raft}
\bibfield{author}{\bibinfo{person}{Diego Ongaro} {and} \bibinfo{person}{John Ousterhout}.} \bibinfo{year}{2014}\natexlab{}.
\newblock \showarticletitle{In Search of an Understandable Consensus Algorithm}. In \bibinfo{booktitle}{\emph{2014 {USENIX} Annual Technical Conference ({USENIX} {ATC} 14)}}. \bibinfo{publisher}{{USENIX} Association}, \bibinfo{pages}{305--319}.
\newblock


\bibitem[{OpenZeppelin}(2021)]%
        {openzeppelin2021governor}
\bibfield{author}{\bibinfo{person}{{OpenZeppelin}}.} \bibinfo{year}{2021}\natexlab{}.
\newblock \bibinfo{title}{Governor: On-chain Governance}.
\newblock \bibinfo{howpublished}{OpenZeppelin Contracts Documentation}.
\newblock
\urldef\tempurl%
\url{https://docs.openzeppelin.com/contracts/4.x/governance}
\showURL{%
\tempurl}


\bibitem[{Optimism Foundation}(2022)]%
        {optimism2022governance}
\bibfield{author}{\bibinfo{person}{{Optimism Foundation}}.} \bibinfo{year}{2022}\natexlab{}.
\newblock \bibinfo{title}{The Future of {Optimism} Governance}.
\newblock \bibinfo{howpublished}{Blog post, Optimism Collective}.
\newblock
\urldef\tempurl%
\url{https://www.optimism.io/blog/the-future-of-optimism-governance}
\showURL{%
\tempurl}


\bibitem[Poupko et~al\mbox{.}(2021)]%
        {poupko2021building}
\bibfield{author}{\bibinfo{person}{Ouri Poupko}, \bibinfo{person}{Gal Shahaf}, \bibinfo{person}{Ehud Shapiro}, {and} \bibinfo{person}{Nimrod Talmon}.} \bibinfo{year}{2021}\natexlab{}.
\newblock \showarticletitle{Building a sybil-resilient digital community utilizing trust-graph connectivity}.
\newblock \bibinfo{journal}{\emph{IEEE/ACM transactions on networking}} \bibinfo{volume}{29}, \bibinfo{number}{5} (\bibinfo{year}{2021}), \bibinfo{pages}{2215--2227}.
\newblock


\bibitem[Raikwar et~al\mbox{.}(2024)]%
        {raikwar2024sokdag}
\bibfield{author}{\bibinfo{person}{Mayank Raikwar}, \bibinfo{person}{Nikita Polyanskii}, {and} \bibinfo{person}{Sebastian M{\"u}ller}.} \bibinfo{year}{2024}\natexlab{}.
\newblock \showarticletitle{{SoK}: {DAG}-based Consensus Protocols}. In \bibinfo{booktitle}{\emph{2024 {IEEE} International Conference on Blockchain and Cryptocurrency ({ICBC})}}. \bibinfo{publisher}{{IEEE}}, \bibinfo{pages}{1--18}.
\newblock
\urldef\tempurl%
\url{https://doi.org/10.1109/ICBC59979.2024.10634358}
\showDOI{\tempurl}


\bibitem[Royo et~al\mbox{.}(2020)]%
        {royo2020decidemadrid}
\bibfield{author}{\bibinfo{person}{Sonia Royo}, \bibinfo{person}{Vicente Pina}, {and} \bibinfo{person}{Jaime Garcia-Rayado}.} \bibinfo{year}{2020}\natexlab{}.
\newblock \showarticletitle{Decide {Madrid}: A Critical Analysis of an Award-Winning e-Participation Initiative}.
\newblock \bibinfo{journal}{\emph{Sustainability}} \bibinfo{volume}{12}, \bibinfo{number}{4} (\bibinfo{year}{2020}), \bibinfo{pages}{1674}.
\newblock
\urldef\tempurl%
\url{https://doi.org/10.3390/su12041674}
\showDOI{\tempurl}


\bibitem[Schneider(2022)]%
        {schneider2022cryptoeconomics}
\bibfield{author}{\bibinfo{person}{Nathan Schneider}.} \bibinfo{year}{2022}\natexlab{}.
\newblock \bibinfo{title}{Cryptoeconomics as a Limitation on Governance}.
\newblock \bibinfo{howpublished}{Mirror}.
\newblock
\urldef\tempurl%
\url{https://ntnsndr.mirror.xyz/zO27EOn9P_62jVlautpZD5hHB7ycf3Cfc2N6byz6DOk}
\showURL{%
\tempurl}
\newblock
\shownote{Working paper}.


\bibitem[Schueffel(2021)]%
        {schueffel2021defi}
\bibfield{author}{\bibinfo{person}{Patrick Schueffel}.} \bibinfo{year}{2021}\natexlab{}.
\newblock \showarticletitle{Defi: Decentralized finance-an introduction and overview}.
\newblock \bibinfo{journal}{\emph{Journal of Innovation Management}} \bibinfo{volume}{9}, \bibinfo{number}{3} (\bibinfo{year}{2021}), \bibinfo{pages}{I--XI}.
\newblock


\bibitem[Shahaf et~al\mbox{.}(2019)]%
        {shahaf2019sybil}
\bibfield{author}{\bibinfo{person}{Gal Shahaf}, \bibinfo{person}{Ehud Shapiro}, {and} \bibinfo{person}{Nimrod Talmon}.} \bibinfo{year}{2019}\natexlab{}.
\newblock \showarticletitle{Sybil-resilient reality-aware social choice}. In \bibinfo{booktitle}{\emph{Proceedings of the 28th International Joint Conference on Artificial Intelligence}}. \bibinfo{pages}{572--579}.
\newblock


\bibitem[Shahaf et~al\mbox{.}(2020)]%
        {shahaf2020genuine}
\bibfield{author}{\bibinfo{person}{Gal Shahaf}, \bibinfo{person}{Ehud Shapiro}, {and} \bibinfo{person}{Nimrod Talmon}.} \bibinfo{year}{2020}\natexlab{}.
\newblock \showarticletitle{Genuine Personal Identifiers and Mutual Sureties for Sybil-Resilient Community Growth}. In \bibinfo{booktitle}{\emph{International Conference on Social Informatics}}. Springer, \bibinfo{publisher}{Springer}, \bibinfo{address}{EU}, \bibinfo{pages}{320--332}.
\newblock


\bibitem[Shapiro(2023a)]%
        {shapiro2023grassrootsBA}
\bibfield{author}{\bibinfo{person}{Ehud Shapiro}.} \bibinfo{year}{2023}\natexlab{a}.
\newblock \showarticletitle{Grassroots Distributed Systems: Concept, Examples, Implementation and Applications (Brief Announcement)}. In \bibinfo{booktitle}{\emph{37th International Symposium on Distributed Computing (DISC 2023). (Extended version: arXiv:2301.04391)}}. \bibinfo{publisher}{LIPICS}, \bibinfo{address}{Italy}, \bibinfo{pages}{47:1, 47:7}.
\newblock


\bibitem[Shapiro(2023b)]%
        {shapiro2023gsn}
\bibfield{author}{\bibinfo{person}{Ehud Shapiro}.} \bibinfo{year}{2023}\natexlab{b}.
\newblock \showarticletitle{Grassroots Social Networking: Serverless, Permissionless Protocols for Twitter/LinkedIn/WhatsApp}. In \bibinfo{booktitle}{\emph{OASIS ’23}} (Rome, Italy). \bibinfo{publisher}{Association for Computing Machinery}.
\newblock
\showISBNx{979-8-4007-0225-9/23/09}
\urldef\tempurl%
\url{https://doi.org/10.1145/3599696.3612898}
\showDOI{\tempurl}


\bibitem[Shapiro(2024)]%
        {shapiro2024gc}
\bibfield{author}{\bibinfo{person}{Ehud Shapiro}.} \bibinfo{year}{2024}\natexlab{}.
\newblock \showarticletitle{Grassroots Currencies: Foundations for Grassroots Digital Economies}.
\newblock \bibinfo{journal}{\emph{arXiv preprint arXiv:2202.05619}} (\bibinfo{year}{2024}).
\newblock


\bibitem[Shapiro(2026)]%
        {shapiro2025atomic}
\bibfield{author}{\bibinfo{person}{Ehud Shapiro}.} \bibinfo{year}{2026}\natexlab{}.
\newblock \showarticletitle{Grassroots Platforms with Atomic Transactions: Social Graphs, Cryptocurrencies, and Democratic Federations}. In \bibinfo{booktitle}{\emph{Proceedings of the 27th International Conference on Distributed Computing and Networking}}. \bibinfo{pages}{71--81, arXiv preprint arXiv:2502.11299}.
\newblock
\urldef\tempurl%
\url{https://doi.org/10.1145.3772309}
\showDOI{\tempurl}


\bibitem[Shapiro and Talmon(2017)]%
        {shapiro2017reality}
\bibfield{author}{\bibinfo{person}{Ehud Shapiro} {and} \bibinfo{person}{Nimrod Talmon}.} \bibinfo{year}{2017}\natexlab{}.
\newblock \showarticletitle{Reality-aware social choice}.
\newblock \bibinfo{journal}{\emph{arXiv preprint arXiv:1710.10117}} (\bibinfo{year}{2017}).
\newblock


\bibitem[Shapiro and Talmon(2025)]%
        {shapiro2025GF}
\bibfield{author}{\bibinfo{person}{Ehud Shapiro} {and} \bibinfo{person}{Nimrod Talmon}.} \bibinfo{year}{2025}\natexlab{}.
\newblock \showarticletitle{Grassroots Federation: Fair Governance of Large-Scale, Decentralized, Sovereign Digital Communities}.
\newblock \bibinfo{journal}{\emph{arXiv preprint arXiv:2505.02208}} (\bibinfo{year}{2025}).
\newblock


\bibitem[Siddarth et~al\mbox{.}(2020)]%
        {siddarth2020watches}
\bibfield{author}{\bibinfo{person}{Divya Siddarth}, \bibinfo{person}{Sergey Ivliev}, \bibinfo{person}{Santiago Siri}, {and} \bibinfo{person}{Paula Berman}.} \bibinfo{year}{2020}\natexlab{}.
\newblock \showarticletitle{Who Watches the Watchmen? {A} Review of Subjective Approaches for Sybil-Resistance in Proof of Personhood Protocols}.
\newblock \bibinfo{journal}{\emph{Frontiers in Blockchain}}  \bibinfo{volume}{3} (\bibinfo{year}{2020}), \bibinfo{pages}{590171}.
\newblock
\urldef\tempurl%
\url{https://doi.org/10.3389/fbloc.2020.590171}
\showDOI{\tempurl}


\bibitem[Spiegelman et~al\mbox{.}(2017)]%
        {spiegelman2017dynamic}
\bibfield{author}{\bibinfo{person}{Alexander Spiegelman}, \bibinfo{person}{Idit Keidar}, {and} \bibinfo{person}{Dahlia Malkhi}.} \bibinfo{year}{2017}\natexlab{}.
\newblock \showarticletitle{Dynamic Reconfiguration: Abstraction and Optimal Asynchronous Solution}. In \bibinfo{booktitle}{\emph{31st International Symposium on Distributed Computing ({DISC} 2017)}} \emph{(\bibinfo{series}{Leibniz International Proceedings in Informatics (LIPIcs)}, Vol.~\bibinfo{volume}{91})}. \bibinfo{publisher}{Schloss Dagstuhl -- Leibniz-Zentrum f{\"u}r Informatik}, \bibinfo{pages}{40:1--40:15}.
\newblock
\urldef\tempurl%
\url{https://doi.org/10.4230/LIPIcs.DISC.2017.40}
\showDOI{\tempurl}


\bibitem[Talmon(2023)]%
        {talmon2023social}
\bibfield{author}{\bibinfo{person}{Nimrod Talmon}.} \bibinfo{year}{2023}\natexlab{}.
\newblock \showarticletitle{Social Choice for {DAOs}}. In \bibinfo{booktitle}{\emph{Proceedings of the 2023 International Conference on Autonomous Agents and Multiagent Systems ({AAMAS} '23)}}. \bibinfo{publisher}{IFAAMAS}.
\newblock


\bibitem[Vizier and Gramoli(2020)]%
        {Vizier2020ComChain}
\bibfield{author}{\bibinfo{person}{Guillaume Vizier} {and} \bibinfo{person}{Vincent Gramoli}.} \bibinfo{year}{2020}\natexlab{}.
\newblock \showarticletitle{{ComChain}: A blockchain with {B}yzantine fault-tolerant reconfiguration}.
\newblock \bibinfo{journal}{\emph{Concurrency and Computation: Practice and Experience}} \bibinfo{volume}{32}, \bibinfo{number}{12} (\bibinfo{year}{2020}), \bibinfo{pages}{e5494}.
\newblock
\urldef\tempurl%
\url{https://doi.org/10.1002/cpe.5494}
\showDOI{\tempurl}


\bibitem[Wang et~al\mbox{.}(2021)]%
        {wang2021non}
\bibfield{author}{\bibinfo{person}{Qin Wang}, \bibinfo{person}{Rujia Li}, \bibinfo{person}{Qi Wang}, {and} \bibinfo{person}{Shiping Chen}.} \bibinfo{year}{2021}\natexlab{}.
\newblock \showarticletitle{Non-fungible token (NFT): Overview, evaluation, opportunities and challenges}.
\newblock \bibinfo{journal}{\emph{arXiv preprint arXiv:2105.07447}} (\bibinfo{year}{2021}).
\newblock


\bibitem[Wood(2016)]%
        {wood2016polkadot}
\bibfield{author}{\bibinfo{person}{Gavin Wood}.} \bibinfo{year}{2016}\natexlab{}.
\newblock \bibinfo{booktitle}{\emph{Polkadot: Vision for a Heterogeneous Multi-Chain Framework}}.
\newblock \bibinfo{type}{White Paper}. \bibinfo{institution}{Web3 Foundation}.
\newblock
\urldef\tempurl%
\url{https://polkadot.com/papers/Polkadot-whitepaper.pdf}
\showURL{%
\tempurl}


\bibitem[Yin et~al\mbox{.}(2019)]%
        {yin2019hotstuff}
\bibfield{author}{\bibinfo{person}{Maofan Yin}, \bibinfo{person}{Dahlia Malkhi}, \bibinfo{person}{Michael~K Reiter}, \bibinfo{person}{Guy~Golan Gueta}, {and} \bibinfo{person}{Ittai Abraham}.} \bibinfo{year}{2019}\natexlab{}.
\newblock \showarticletitle{HotStuff: BFT consensus with linearity and responsiveness}. In \bibinfo{booktitle}{\emph{Proc. ACM PODC'19}}. \bibinfo{pages}{347--356}.
\newblock


\end{thebibliography}
